\documentclass[aps,prd,twocolumn,nofootinbib,preprintnumbers,showkeys,superscriptaddress]{revtex4-2}

\usepackage{newtxtext}
\usepackage{newtxmath}
\usepackage{mathrsfs}

\usepackage{graphicx}
\usepackage{bm}
\usepackage{amsmath}
\usepackage{float}
\usepackage{multirow}
\usepackage{slashed}
\usepackage{xcolor}
\usepackage{mathtools,braket}
\usepackage{color,physics}

\def\orcid#1{\kern .08em\href{https://orcid.org/#1}{\includegraphics[keepaspectratio,width=0.7em]{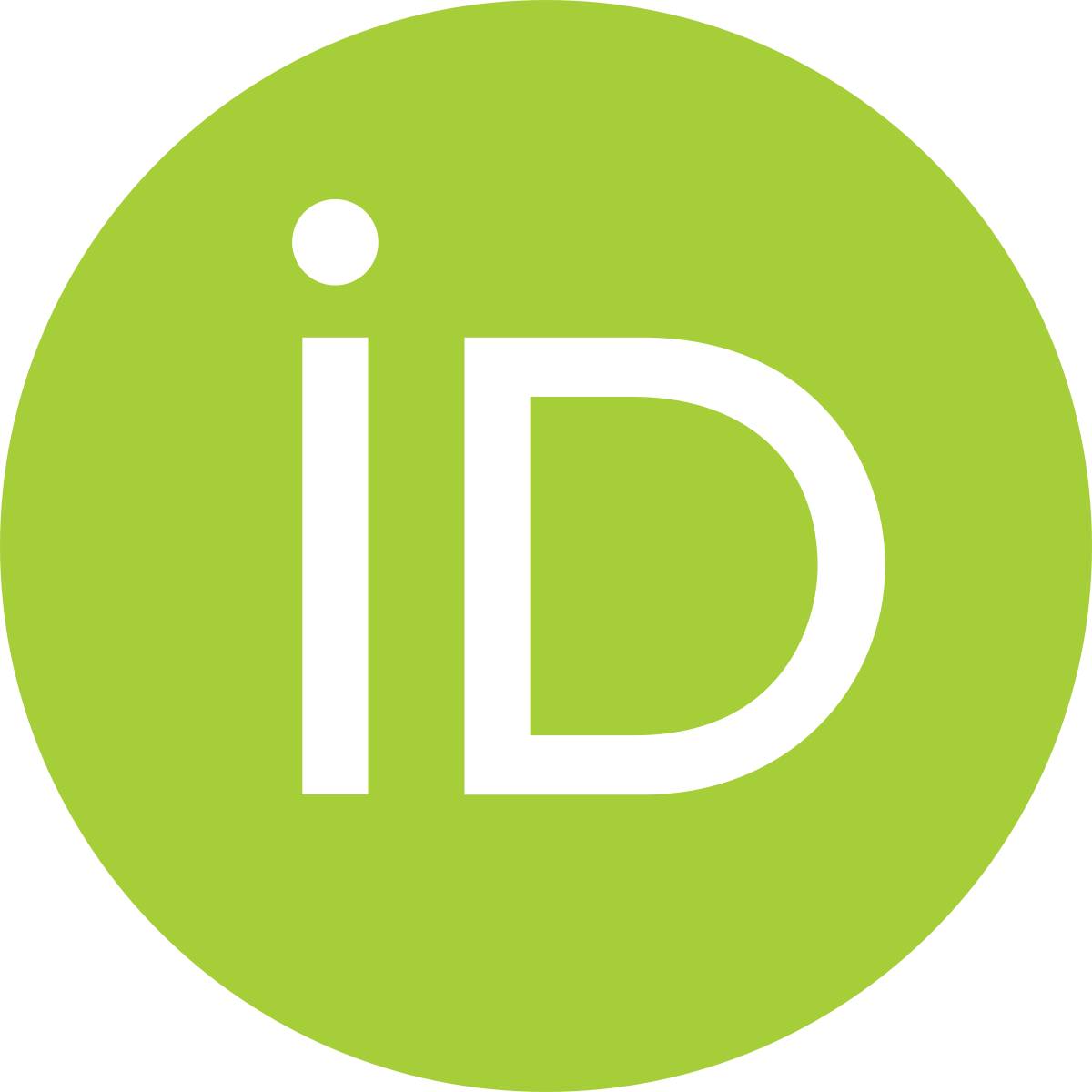}}}

\usepackage[colorlinks=true, pdfstartview=FitV, bookmarks=true, bookmarksnumbered=true, breaklinks]{hyperref}
\hypersetup{linkcolor=blue, citecolor=blue, urlcolor=blue}
\newcommand{\hrefr}[2]{\href{#2}{#1}}

\begin{document}
\title{Strong decays of multi-strangeness baryon resonances in the quark model}

\author{Ahmad Jafar Arifi\orcid{0000-0002-9530-8993}}
\email{ahmad.jafar.arifi@apctp.org}
\affiliation{Asia Pacific Center for Theoretical Physics, Pohang, Gyeongbuk 37673, Korea}

\author{Daiki Suenaga\orcid{0000-0002-6928-4123}}
\email{suenaga@rcnp.osaka-u.ac.jp}
\affiliation{Research Center for Nuclear Physics, Osaka University, Ibaraki, Osaka 567-0047, Japan}

\author{Atsushi Hosaka\orcid{0000-0003-3623-6667} }
\email{hosaka@rcnp.osaka-u.ac.jp}
\affiliation{Research Center for Nuclear Physics, Osaka University, Ibaraki, Osaka 567-0047, Japan}
\affiliation{Advanced Science Research Center, Japan Atomic Energy Agency, Tokai, Ibaraki 319-1195, Japan}

\author{Yongseok Oh\orcid{0000-0001-9822-8975} }
\email{yohphy@knu.ac.kr}
\affiliation{Department of Physics, Kyungpook National University, Daegu 41566, Korea}
\affiliation{Asia Pacific Center for Theoretical Physics, Pohang, Gyeongbuk 37673, Korea}

\date{\today}

\begin{abstract}
Decay properties of multi-strangeness $\Xi$ and $\Omega$ baryon resonances are investigated within the constituent quark model by 
including relativistic corrections.  
The strong decay widths of various $\Xi$ and $\Omega$ resonances are computed and tested for internal quark configurations to pin down 
their spin-parity ($J^P$) quantum numbers and internal structures. 
We found that the Roper-like resonances in the multi-strangeness sector have large decay widths bearing resemblances to their siblings 
in light and heavy baryon sectors.
In addition, we obtained the decay ratio for the $\Omega(2012)$ as $\Gamma{(\Omega\to \Xi \bar{K} \pi)} / \Gamma{(\Omega\to \Xi \bar{K})}=4.5\%$, 
which is consistent with the experimental data when $J^P=3/2^-$ is assigned.
This observation holds the possibility that the $\Omega(2012)$ could naturally be explained as a three-quark state in the quark model.
The decays of other low-lying $\Xi$ and $\Omega$ resonances are systematically analyzed, which would be useful to unveil the structure 
of these resonances in future experiments.   

\end{abstract}

\maketitle

\section{Introduction}

As pointed out in the recent issue of the Particle Data Group (PDG)~\cite{PDG20}, most information on the $\Xi$ and $\Omega$ 
baryons was extracted from the bubble chamber experiments in the 1980s and no significant information on these baryons 
has been added for a few decades.
However, the discoveries of $\Xi$ and $\Omega$ resonances at current accelerator facilities by the Belle, BESIII, and LHCb
Collaborations~\cite{BESIII-15b,Belle-18,Belle-18b,Belle-19,BESIII-19,LHCb-20,Belle-21} changed the situation and triggered
the recent interests in multi-strangeness baryons.
Furthermore, experimental studies on multi-strangeness baryons are discussed or planned at next generation 
facilities~\cite{KLF-20,PANDA-21,JPARC-21}.

In understanding the strong interactions, multi-strangeness baryons offer a unique window showing some interesting aspects. 
For example, the $\Omega$ baryons, containing only strange valence quarks, would provide a good tool to test 
effective models of baryons.
In this case, the pion cloud contribution is expected to be suppressed as there is no valence up/down quarks inside. 
Although the kaon cloud effects may not be excluded~\cite{JKKS21}, $\Omega$ baryons allow more direct access to valence quark 
dynamics in the core region of baryons, while it is very unlikely in the case of light baryons. 
In this respect, the radial excitation of the $\Omega$ baryon is of great interest as compared to the Roper resonance $N(1440)$ case 
where the pion clouds are expected to play an important role~\cite{SJKLMS09,BR17}.
Another issue is the structure of the newly discovered $\Omega(2012)$~\cite{Belle-18}. 
Its structure is yet to be resolved and it is under debate whether it is a hadronic molecule~\cite{PO18,GL19,ITO20,Valderrama18,LZWXG20} 
or a standard quark model state~\cite{XZ18,LWLZ19,WGLXZ18}.

Having two strange quarks and one light up/down quark, the $\Xi$ baryon is theoretically challenging.
The strange quark mass is not large enough to respect the heavy quark symmetry, but it is not small enough to apply the SU(3) 
flavor symmetry.
Therefore, it offers a way to investigate the mass dependence, or SU(3) flavor symmetry breaking effects in baryon structure being 
complementary to $S=-1$ hyperons.
There have been many theoretical attempts to understand $\Xi$ baryon spectrum~\cite{HN20}, not only from the quark model~\cite{CI86}, 
but also from other theoretical models such as the Skyrme model~\cite{Oh07}, chiral unitary model~\cite{ROB02,GLN03,SOV04}, and 
QCD sum rules approach~\cite{AAS18}, etc.
The predicted $\Xi$ spectra show strong model dependence as in the case of $\Omega$ baryons.
In particular, some peculiar structures of $S=-1$ hyperon resonances or nucleon resonances are expected to be repeated 
in multi-strangeness baryon resonances~\cite{Oh11}.
Therefore, investigating such resonances will help unveil the hadron structure.

\begin{figure}[t]
	\centering
	\includegraphics[scale=0.43]{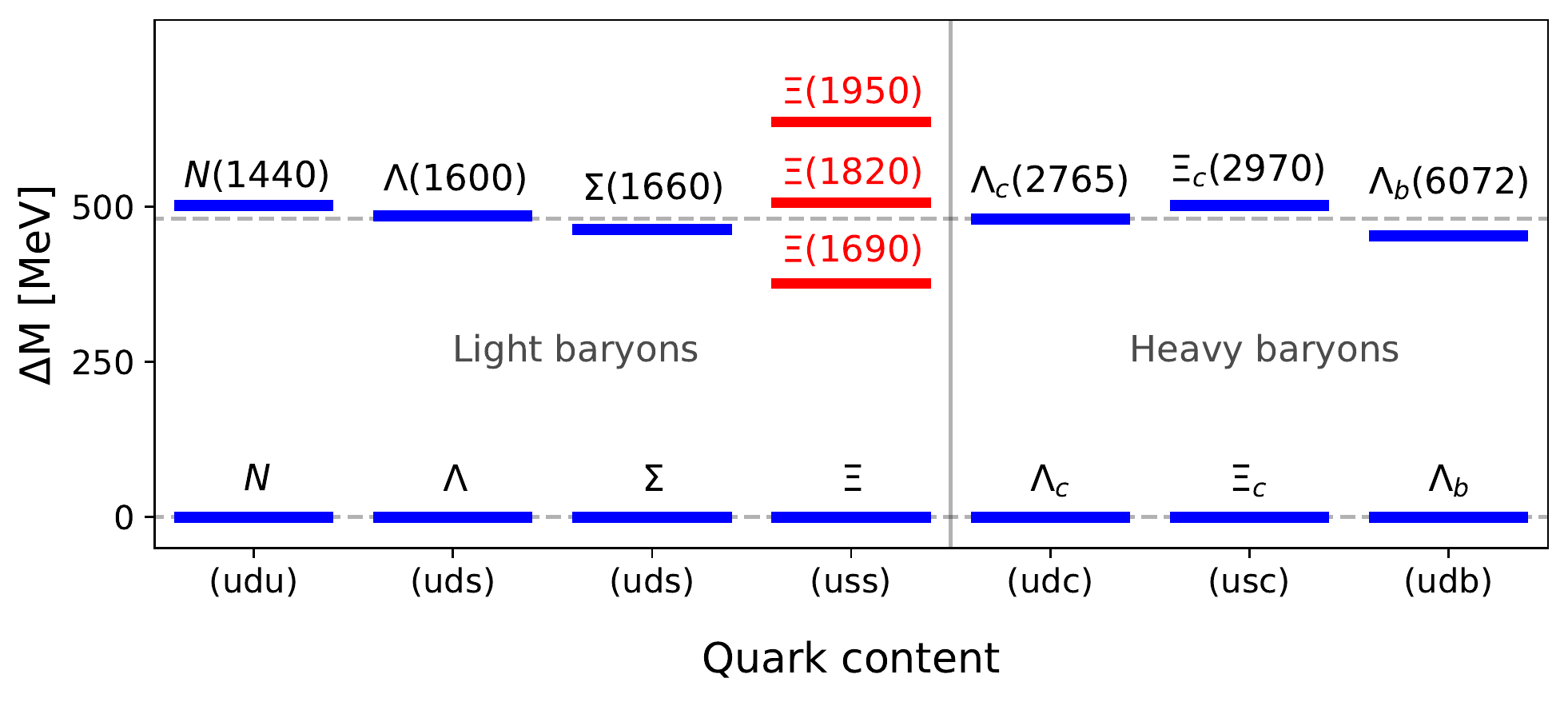}
	\caption{\label{roper} 
	Roper-like resonances with various quark flavor contents. 
		The red bands are the observed $\Xi$ resonances listed in the Particle Data Group (PDG)~\cite{PDG20} 
		with excitation energy of around 500~MeV, but none of them are yet identified as the radially excited state with $J^P=1/2^+$.
		In particular, the $\Xi(1820)$ is identified to have $J^P = 3/2^-$~\cite{PDG20}, and is excluded from the candidate.
		We also note that the spin-parity of the $\Xi(1690)$ was claimed to be $J^P = 1/2^-$ by the 
		BaBar Collaboration~\cite{BABAR08}.}
\end{figure}

One of the major subjects to be studied in the present work is the analogous states to the Roper resonance.  
In spite of recent experimental and theoretical efforts, the first radial excitations of the $\Xi$ and $\Omega$ baryons, which are analogous to 
the Roper resonance, are yet to be identified. 
On the other hand, the putative Roper-like resonances with heavy-quark flavor are recently reported in the experiments at
current experimental facilities~\cite{Belle-20b,LHCb-20c}.%
\footnote{Since the Roper resonance is expected to be the first radial excitation of the ground state in the quark model, its quantum
numbers should be $J^P = 1/2^+$. 
The recent observation of the $\Xi_c(2970)$~\cite{Belle-20b} and the $\Lambda_b(6072)$~\cite{LHCb-20c} indicates that these baryons 
may be the analog states of the Roper resonance.}
Interestingly, they have similar excitation energies of around 500~MeV from their respective ground states~\cite{ANHT20b}.
Consequently, as shown in Fig.~\ref{roper}, one may expect that there exists a Roper-like $\Xi$ resonance at that energy range.
A similar resonance may exist for the $\Omega$ baryon spectrum.
In this paper, we explore this possibility as well.

The Roper resonance is understood as the first radially excited state of the nucleon and, therefore, has $J^P = 1/2^+$.
Since its first discovery~\cite{Roper64,AFR85}, a lot of investigations have been performed and it is now widely accepted that
the radially excited three-quark core is dressed by pion clouds which have a nontrivial role~\cite{SJKLMS09,BR17,FH21}.
Thus, finding its missing siblings and understanding the similarities and differences in multi-strangeness hyperons would shed light 
on our understanding of the structures of, in particular, radially excited baryon resonances.

In the literature, there are numerous theoretical predictions for radially excited states in multi-strangeness baryon spectra, whose results are
highly model dependent.
For $\Xi$ baryons, the masses are predicted to be between 1730 and 1960~MeV 
(or  the excitation energy $\Delta M \approx 410$ to 650~MeV). 
Similarly, for $\Omega$ baryons, the masses are predicted to be between 2060 and 2180 MeV 
(or  $\Delta M \approx 390$ to 510~MeV). 
We refer the details to Refs.~\cite{CIK81,CI86,BIL00,LMP01b,PR07,MPS08,CM09,LWLZ19,CI86,PR07,SF14,FG15,YHHOS15, MR21,MR21b}.
In general, the above model calculations indicate that the predicted mass gap between the Roper-like strange resonances and their respective 
ground states is around $500$~MeV as mentioned earlier.
However, since there could be many resonances in this mass region, it is very important to identify the spin-parity quantum numbers of the 
resonances and decay patterns for understanding their structures.
The spin-parity quantum numbers of resonances are usually determined by the angular distributions of their decay products or 
by measuring polarization observables in the production processes.
The purpose of the present work is to investigate the spin-parity quantum numbers of multi-strangeness resonances through their decay processes.
In particular, we put more emphasis on the Roper-like resonances that would be observed at the present experimental facilities 
in the near future.
As we have discussed in our previous work~\cite{ASH21}, the relativistic effects play a crucial role in the decays of Roper-like 
resonances and we further explore their role in the present work.

This article is organized as follows.
In Sec.~\ref{sec:wf}, we briefly review the baryon wave functions in the quark model.
The basics of our calculations and model parameters are introduced as well.
Then, in Sec.~\ref{sec:QMC}, we discuss the quark-meson interactions with the relativistic effects that will be used in this work
for studying the decays of hyperon resonances.
Our results on the decays of $\Xi$ and $\Omega$ resonances are presented and discussed in Sec.~\ref{sec:Xi} and 
Sec.~\ref{sec:Omega}, respectively.
Section~\ref{sec:summary} summarizes the present work.
In Appendix, we explain how to compute the three-body decay of the $\Omega(2012)$.

\section{Baryon wave functions} \label{sec:wf}

In this section, we briefly review the wave functions of baryons that will be used to estimate their strong decays in the quark model.

\subsection{Spatial wave function}

The Hamiltonian of the nonrelativistic constituent quark model reads
\begin{eqnarray}
	\mathcal{H} = -\sum_{i=1}^3 \frac{\boldsymbol{\nabla}^2_i}{2m_i} 
	+ \frac{k}{2} \sum_{i<j} \abs{ \bm{r}_j^{} -\bm{r}_i^{} }^2, 
\label{eq:Hamiltonian}
\end{eqnarray}
where $m_i$ and $\bm{r}_i^{}$ are the mass and the coordinate of the $i$-th constituent quark inside a baryon, respectively, 
and we assume the flavor-independent harmonic oscillator potential with the same spring constant $k$ for inter-quark interactions.
Although the strange quark mass is larger than that of the light up/down quarks, in the present exploratory 
computations, we use the averaged quark mass respecting the SU(3) flavor symmetry.

\begin{table*}[t]
	\begin{ruledtabular}
		\renewcommand{\arraystretch}{1.6}
		\centering
		\caption{\label{total} Wave functions of baryons in the $\mbox{SU(6)} \times \mbox{O(3)}$ group structure.
			The Clebsch-Gordan coefficients for the addition of spin and orbital angular momenta are omitted.
			The spatial, spin, and flavor wave functions are denoted by  $\Psi$, $\chi$ and $\phi$, respectively.   }
		\begin{tabular}{ccccccc}
			State									& 	$L$	& 	$S$ 			&	 $J^P$ 			& Wave function 	\\ \hline
			$N=0$							\\ \hline
			$\ket{56, {}^{2}8,0,0,\tfrac{1}{2}^+}$		& 	0	&	$\frac{1}{2}$	&	$\frac{1}{2}^+$	& $\frac{1}{\sqrt{2}}\Psi^S_{000} (\phi_\rho \chi_\rho + \phi_\lambda \chi_\lambda)$ \\ 
			$\ket{56, {}^{4}10,0,0,\tfrac{3}{2}^+}$		&	0	&  	$\frac{3}{2}$	&	$\frac{3}{2}^+$	& $\Psi^S_{000} ~(\phi_S \chi_S)$  \\ \hline
			$N=1$ 								\\ \hline
			$\ket{70, {}^{2}10,1,1,J^-}$		& 1	&  	$\frac{1}{2}$	&	$\frac{1}{2}^-, \frac{3}{2}^-$	& $	\frac{1}{\sqrt{2}}\left(  \Psi^\lambda_{11m} \left( \phi_S  \chi_\lambda\right)+  \Psi^\rho_{11m} \left( \phi_S \chi_\rho\right) \right) $   \\ 
			$\ket{ 70, {}^{2}8,1,1,J^-}$		&  1	&	$\frac{1}{2}$	&	$\frac{1}{2}^-,\frac{3}{2}^-$	& 
			$ \frac{1}{2} \Psi^\lambda_{11m}  \left(\phi_\rho \chi_\rho- \phi_\lambda \chi_\lambda \right)+\frac{1}{2}  \Psi^\rho_{11m} \left(\phi_\rho \chi_\lambda+ \phi_\lambda \chi_\rho\right)$\\
			$\ket{70, {}^{4}8,1,1,J^-}$		& 	1	&  	$\frac{3}{2}$	&	$\frac{1}{2}^-,\frac{3}{2}^-,\frac{5}{2}^-$	& $ \frac{1}{\sqrt{2}}\left( \Psi^\lambda_{11m} \left(\phi_\lambda \chi_S \right) +\Psi^\rho_{11m} \left(\phi_\rho\chi_S \right) \right) $\\  \hline
			$N=2$ 									\\ \hline
			$\ket{56, {}^{2}8,2,0,\tfrac{1}{2}^+}$		& 	0	&	$\frac{1}{2}$	&	$\frac{1}{2}^+$	& $\frac{1}{\sqrt{2}} \Psi^S_{200} (\phi_\rho \chi_\rho + \phi_\lambda \chi_\lambda) $ \\ 
			$\ket{ 56, {}^{4}10,2,0,\tfrac{3}{2}^+}$		& 	0	&  	$\frac{3}{2}$	&	$\frac{3}{2}^+$	& $\Psi^S_{200} (\phi_S \chi_S)$  \\
		\end{tabular}
		\renewcommand{\arraystretch}{1}
	\end{ruledtabular}
\end{table*}

For a three-body system of the equal mass, the Jacobi coordinates are defined by
\begin{subequations}
\begin{eqnarray}
	\boldsymbol{\rho} &=&  \bm{r}_1^{} - \bm{r}_2^{},\\
	\boldsymbol{\lambda} &=& \frac{1}{2} \left( \bm{r}_1^{} + \bm{r}_2^{}\right) - \mathbf{r}_3,\\
	\boldsymbol{R} &=& \frac{1}{3} \left( \bm{r}_1^{} + \bm{r}_2^{} +\bm{r}_3^{} \right).
\end{eqnarray}
\end{subequations}
The harmonic oscillator Hamiltonian of Eq.~(\ref{eq:Hamiltonian}) can then be written in terms of the Jacobi coordinates as
\begin{eqnarray}
	\mathcal{H} &=&  -\frac{\boldsymbol{\nabla}^2_{R}}{2M} - \frac{\boldsymbol{\nabla}^2_{\rho}}{2m_\rho}   
	+ \frac{m_\rho \omega_\rho^2}{2} \boldsymbol{\rho}^2  
	- \frac{\boldsymbol{\nabla}^2_{\lambda} }{2m_\lambda}
	+ \frac{m_\lambda \omega_\lambda^2}{2} \boldsymbol{\lambda} ^2,
\label{eq:Hamiltonian2}
\end{eqnarray}
where the total mass is $M=3m$ and the reduced masses are given by
\begin{eqnarray}
	m_\lambda = \frac{2m}{3},     \qquad
	m_\rho = \frac{m}{2}
\end{eqnarray}
with $m$ being the averaged quark mass inside the baryon.
The harmonic oscillator energy of $\lambda$- and $\rho$-modes are degenerate as we use the averaged quark mass, which leads to
\begin{eqnarray}
	\omega_\lambda = \omega_\rho = \sqrt{\frac{3k}{m}}.
\end{eqnarray}
Their corresponding range parameters are, therefore, written as
\begin{eqnarray}
	a_\lambda = \sqrt{m_\lambda \omega_\lambda},    \qquad
	a_\rho = \sqrt{m_\rho \omega_\rho},
\end{eqnarray}
which then leads to the relation,
\begin{eqnarray}
	a^2_\rho =\frac{3}{4} a^2_\lambda.
\end{eqnarray}
Furthermore, each quark momentum can be written in terms of the Jacobi momenta, $\bm{P}$,
$\bm{p}_\lambda^{}$, and $\bm{p}_\rho^{}$, which are conjugate to the position vectors, 
$\bm{R}$, $\bm{\lambda}$, and $\bm{\rho}$, respectively. 
Explicitly, we have
\begin{subequations}
\begin{eqnarray}
	\bm{p}_1^{} &=& \frac{1}{3} \bm{P}  + \frac{1}{2} \bm{p}_\lambda^{} - \bm{p}_\rho^{},\\
	\bm{p}_2^{} &=&  \frac{1}{3} \bm{P}  + \frac{1}{2} \bm{p}_\lambda^{} + \bm{p}_\rho^{},\\
	\bm{p}_3^{} &=& \frac{1}{3}  \bm{P}  -  \bm{p}_\lambda^{}.
\end{eqnarray}
\end{subequations}

Because of the separable form of the Hamiltonian~(\ref{eq:Hamiltonian2}), the spatial wave function of the three-quark state 
is expressed by using the separation of variables as
\begin{eqnarray}
	\Psi (\bm{r}_1^{}, \bm{r}_2^{}, \bm{r}_3^{} ) = 
	\psi^{(\lambda)} (\boldsymbol{\lambda}) \psi^{(\rho)} (\boldsymbol{\rho}) e^{i \boldsymbol{P} \cdot \boldsymbol{R}},
\end{eqnarray}
where $\psi^{(\lambda)} (\boldsymbol{\lambda})$ and $\psi^{(\rho)} (\boldsymbol{\rho})$ are the harmonic oscillator wave functions in 
the Jacobi coordinates.
Each spatial wave function is again written as
\begin{eqnarray}
	\psi_{nlm} (\boldsymbol{r}) = R_{nl} (r) Y_{lm}(\hat{r}),
\end{eqnarray}
where $R_{nl} (r)$ is the radial wave function and $Y_{lm}$ is the spherical harmonics with $\bm{r}$ standing for $\bm{\lambda}$ and $\bm{\rho}$.

\subsection{Spin-flavor and total wave functions}

Following the SU(3) flavor symmetry, baryons made of three quarks are classified into flavor multiplets as
\begin{eqnarray}
	\mathbf{3} \otimes \mathbf{3} \otimes \mathbf{3}
	= \mathbf{10}_S \oplus \mathbf{8}_{M} \oplus \mathbf{8}_{M} \oplus \mathbf{1}_{A},
\end{eqnarray}
where the subscripts indicate symmetric $(S)$, mixed-symmetric $(M)$, or antisymmetric $(A)$ states under the interchange of any two quarks.
For the spin part, spin wave functions are classified as 
\begin{eqnarray}
	\mathbf{2} \otimes \mathbf{2} \otimes \mathbf{2}= \mathbf{4}_S \oplus \mathbf{2}_{M} \oplus \mathbf{2}_{M}.
\end{eqnarray}
The spin-flavor wave functions are then described by SU(6) symmetry.
Thus, baryon multiplets are decomposed into
\begin{eqnarray}
	\mathbf{6} \otimes \mathbf{6} \otimes \mathbf{6}= \mathbf{56}_S \oplus \mathbf{70}_{M} \oplus \mathbf{70}_{M} \oplus \mathbf{20}_{A}.
\end{eqnarray}
The standard spin and flavor wave function of baryons can be found, for example, in Ref.~\cite{HT}.

The total wave function of a baryon is then given by
\begin{eqnarray}
	\ket{qqq} = \ket{ \text{color}}_A  \ket{ \text{space, spin, flavor} }_S.
\end{eqnarray}
Since the color part is always anti-symmetric, we need to construct a symmetric combination of spatial, spin, and flavor wave functions. 
The wave function is then denoted by $\ket{ N_6, {}^{2S+1} N_3, N, L, J^P }$, where $N_6$ and $N_3$ correspond to the SU(6) and SU(3) 
representations, respectively. 
The superscript $S$ refers to the total intrinsic spin of the system, and $N$ and $L$ are the principal quantum number and 
orbital angular momentum, respectively.
Finally, the spin and parity quantum numbers of the baryon are denoted by $J^P$.
The total wave functions of baryons up to the first radial excitations obtained in this way are summarized in Table~\ref{total}.

\section{Quark-Meson Coupling Interactions} \label{sec:QMC}

\begin{figure}[t]
	\centering
	\includegraphics[scale=0.2]{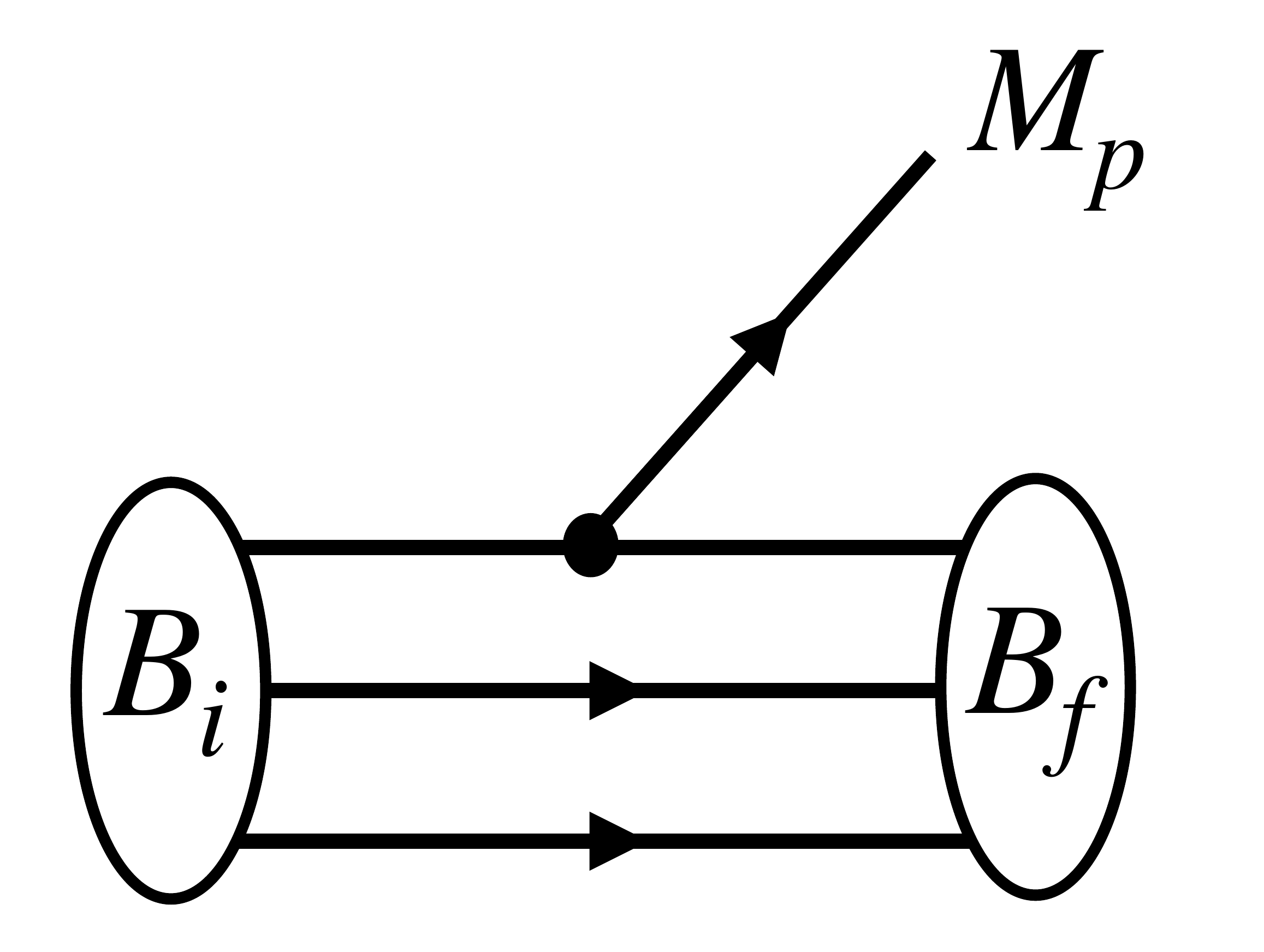}
	\caption{\label{fig:decay} 
	One pseudoscalar meson $(M_p = \pi, K)$ emission decay of a baryon, $B_i(P_i) \to B_f(P_f) + M_p(q)$, in the quark model,
	where $P_i$, $P_f$, and $q$ are the momenta of the corresponding particles.
	The outgoing meson may couple to any quark inside the baryon. }
\end{figure}

The quark model picture adopted in the present work for one-meson emission decay of a baryon, i.e., $B_i(P_i) \to B_f(P_f) + M_p(q)$,
is depicted in Fig.~\ref{fig:decay}.
Here we only consider the decays where $M_p$ corresponds to a pseudoscalar meson.
The interaction between the outgoing pseudoscalar meson and the quark inside the baryon takes the axial-vector type and is written as
\begin{eqnarray}
	\mathscr{L}_{M_p qq} = \frac{g_A^q}{2f_p} \bar{q} \gamma_\mu\gamma_5 \vec{\lambda} \cdot \partial^\mu \vec{M}_p q ,
\label{inter}
\end{eqnarray}
where $g^q_A=1$ is the quark axial-vector coupling constant, $\vec M_p$ is the pseudoscalar meson octet field, and 
$\vec{\lambda}$ is the SU(3) Gell-Mann matrix.
The empirical data for decay constants are $f_\pi=93$ MeV and $f_K=111$ MeV for pion and kaon, respectively.
This interaction is inspired by the low-energy theorem of chiral symmetry and has been widely used in quark model 
calculations~\cite{NYHON16,XZ13,ZZ07}.

In most calculations, the nonrelativistic limit of the interaction (\ref{inter}) has been taken by considering the leading terms up to $O(1/m)$,
which leads to the Hamiltonian as
\begin{eqnarray}
	\mathcal{H}_{\rm NR} = g \left[ \boldsymbol{\sigma}\cdot \boldsymbol{q} 
	+ \frac{\omega}{2m}\left( \boldsymbol{\sigma}\cdot \boldsymbol{q} 
	-2 \boldsymbol{\sigma}\cdot \boldsymbol{p}_i \right) \right],
 \label{nonrel}
\end{eqnarray}
where $m$ is the constituent quark mass and $g = g_A^q/ 2f_p$.
The 4-momentum of the outgoing pseudoscalar meson is $q=(\omega, \boldsymbol{q})$, and the 3-momentum of the quark inside 
the initial state baryon is $\bm{p}_i$.

In order to estimate the relativistic corrections, we perform the Foldy-Wouthuysen transformation for the Lagrangian~(\ref{inter}). 
These correction terms are important, in particular, for the case where the leading terms are suppressed as in the case for the 
decay of the Roper-like states.
This procedure leads to~\cite{ASH21}
\begin{eqnarray}
	\mathcal{H}_{\rm RC} =\frac{g}{8 m^2} \biggl[ m_p^2 \boldsymbol{\sigma}\cdot \boldsymbol{q} 
	+  2 \boldsymbol{\sigma}\cdot (\boldsymbol{q} -2\boldsymbol{p}_i ) \times (\boldsymbol{q} \times \boldsymbol{p}_i)  
	\biggr] ,
\label{relcor}
\end{eqnarray}
as the ${O}(1/m^2)$ corrections. 
Here $m_p$ is the mass of the emitted pseudoscalar meson.
As we have seen in Ref.~\cite{ASH21}, the second term in $\mathcal{H}_{\rm RC}$, which is proportional to $\boldsymbol{p}_i^2$,
has a nontrivial role, in particular, for the Roper-like states.
As an analogy in the electromagnetic interactions, such a term appears as the spin-orbit coupling 
in the relativistic corrections.

The quark model summarized in the previous section has two parameters, namely, the constituent quark mass $m$ and the spring constant $k$.
In this work, we take the average of the mass of the up/down quark (350~MeV) and the strange quark (450~MeV). 
As a result, the constituent quark mass for $\Xi$ baryons is $m = 0.417$~GeV and that for $\Omega$ baryons is
$m = 0.450$~GeV.
The spring constant is fixed to be $k = 0.02$~GeV$^3$ to reproduce baryon radius $\sqrt{\Braket{R^2}}=\sqrt{1/(3a_\lambda^2) + 1/(4a_\rho^2)}$ 
of around 0.5~fm that is obtained from the lattice calculations of Refs.~\cite{AKKNP10,CEOT15}.
We note that the radius computed in this quark model corresponds to the baryon core in which the valence quarks 
carrying the baryon density are localized~\cite{FKW20}.
Collecting all these information, we use $(\omega_\lambda, a_\lambda, a_\rho) = (0.379, 0.325, 0.281)$~GeV for $\Xi$ baryons and 
$(\omega_\lambda, a_\lambda, a_\rho) = (0.365, 0.331, 0.287)$~GeV for $\Omega$ baryons.

\begin{figure}[t]
	\centering
	\includegraphics[scale=0.3]{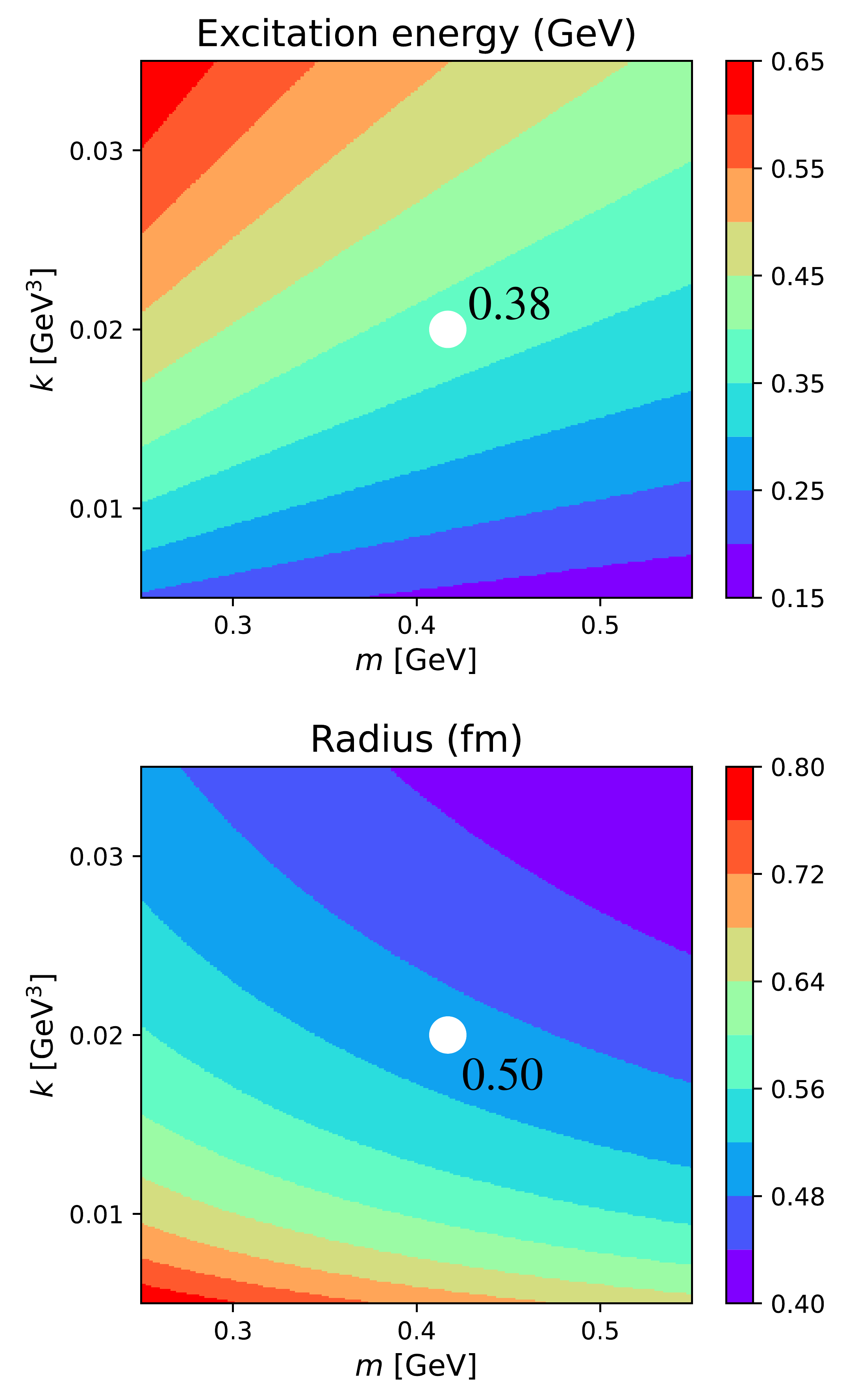}
	\caption{\label{fig:parameter} 
	Contour plot for the excitation energy $(\omega_\lambda,\omega_\rho)$ and radius $\sqrt{\Braket{ R^2 }}$ of $\Xi$ baryon 
	as functions of the quark model parameters $m$ and $k$.
	White dots indicate the natural quark model parameters that reproduce the baryon core radius.}
\end{figure}

However, since the quark model results depend on the model parameters, namely, the quark mass $m$ and the spring constant $k$ 
(or translated to the oscillator parameter $\alpha$~\cite{XZ13}), we will provide contour plots for our numerical results as
functions of $m$ and $k$, which will allow estimating the uncertainties of our results.
This then shows how sensitive the obtained quantity is to the quark model parameters.
As an example, shown in Fig.~\ref{fig:parameter} are the dependence of the excitation energy $\omega_\lambda$ and the radius 
$\sqrt{\Braket{R^2}}$ on the quark model parameters. 
The central values of the fitted parameters are shown as white dots in the figures.

\section{ $\mathbf{\Xi}$ baryon decays} \label{sec:Xi}

In this section, we discuss the decays of $\Xi$ baryons.
Some earlier works would introduce an additional universal suppression parameter $\delta$ for the coupling constant to reproduce the experimental value of the decay width of 
$\Xi(1530)$~\cite{XZ13}.
In the present work, however, we do not introduce any additional parameters and keep $g^q_A=1$.
We shall find that this suppression could be naturally explained by the relativistic corrections.
We will come back to this point later.

The decay width of a two-body decay of $B_i \to B_f M_p$ is expressed in the helicity basis as
\begin{eqnarray}
	\Gamma = \frac{1}{4\pi} \frac{q}{2M_i^2}\frac{1}{2J+1} \sum_h \abs{ A_h }^2,
\end{eqnarray}
where $q= \abs{P_f}$ is the momentum of the outgoing meson, $A_h$ is the helicity amplitude, and $M_i$ is the mass of the initial state
baryon with spin $J$. 
We refer the details, for example, to Ref.~\cite{NYHON16} for representing the decay amplitude in the helicity basis.

\subsection{\boldmath $\Xi(1530)$ resonance}

Before investigating the decays of other excited states, we work on the decay of the $\Xi(1530)$ that belongs to the decuplet 
with $J^P=3/2^+$ for checking the applicability of our model.
The $\Xi(1530)$ mainly decays into $\Xi\pi$ and its total decay width is measured to be $\Gamma=9.1 \pm 0.5$ MeV~\cite{PDG20}.
In the quark model, the state of the $\Xi(1530)$ corresponds to $\ket{ 56, {}^{4}10,0,0,\tfrac{3}{2}^+}$.

\begin{figure}[t]
	\centering
	\includegraphics[scale=0.3]{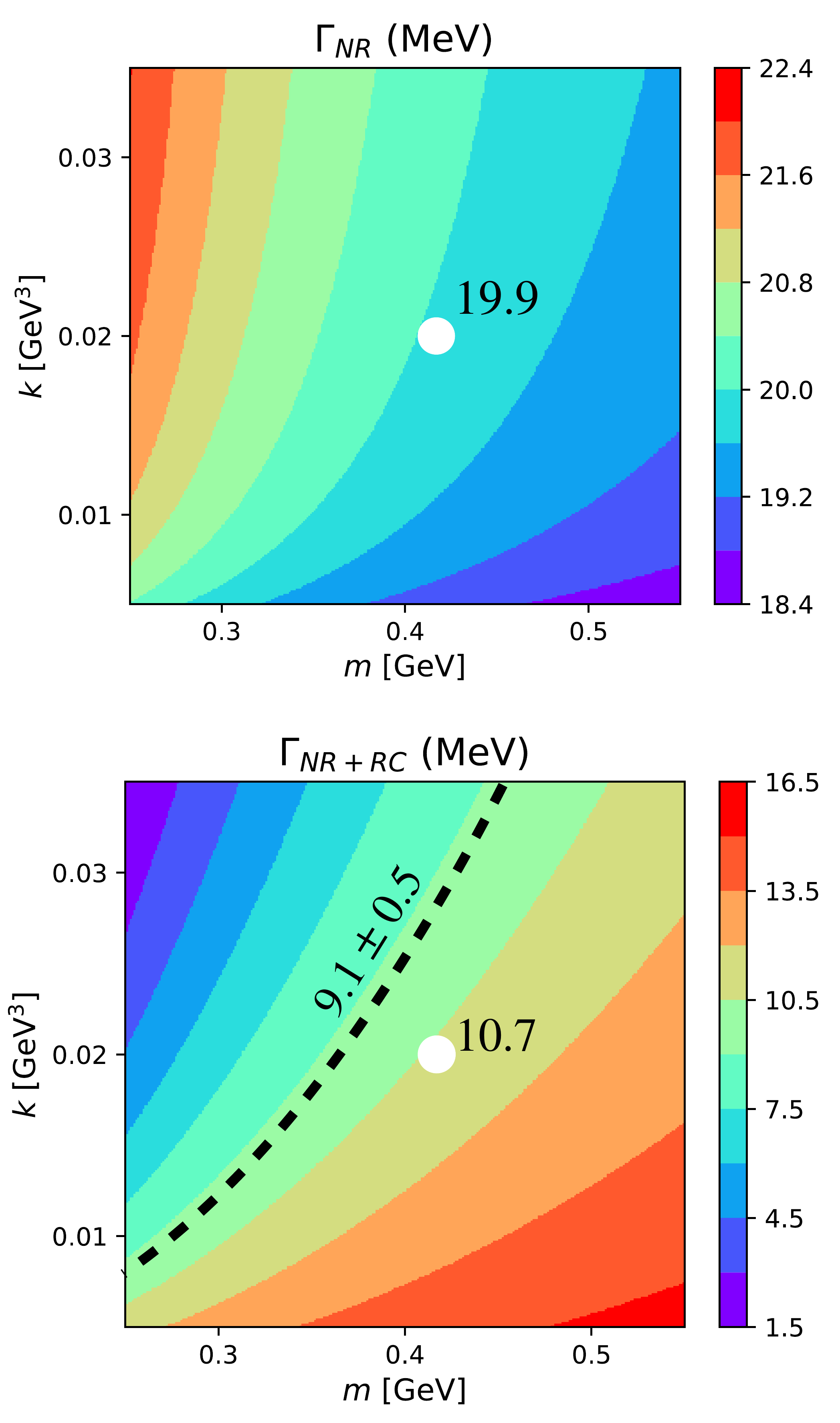}
	\caption{\label{fig:xi1530} 
	Contour plot for the predicted decay width of $\Xi(1530)^0$ as a function of the quark mass $m$ and spring constant $k$. 
	$\Gamma_{\rm NR}$ is the decay width obtained in the nonrelativistic limit, while $\Gamma_{\rm NR + RC}$ is the results including
	the relativistic corrections.
	The black dashed line is the experimental data of $\Gamma = 9.1\pm0.5$~MeV~\cite{PDG20}.}
\end{figure}

Tabulated in Table~\ref{tab:xi1530} are the predicted decay widths of $\Xi(1530)$ and other decuplet baryons in the quark model of the present work.
When the nonrelativistic Hamiltonian~(\ref{nonrel}) is adopted, the decay width of $\Xi(1530)$ is found to be larger than the experimental value 
by a factor of two, which is similar to the observation made in the heavy baryon sector~\cite{ASH21}.
Furthermore, the upper panel of Fig.~\ref{fig:xi1530} shows that it is impossible to reproduce the observed decay width of
$\Gamma \approx 9.1$~MeV within the reasonable range of the model parameters.
In order to reconcile this discrepancy, phenomenological suppression would be called for through the variation of the quark-axial coupling 
constant or the introduction of a universal suppression factor~\cite{XZ13}.
However, we found that the inclusion of the relativistic corrections provides a natural explanation for the suppression and makes the predictions closer
to the observed values.
This can be seen in the lower panel of Fig.~\ref{fig:xi1530} which shows 
that with the inclusion of the relativistic corrections, the decay width, labeled as
$\Gamma_{\rm NR+RC}$, is suppressed and closer to the measured data within the reasonable parameter range.
Our result here also implies that our method works sufficiently well for the $\Xi$ baryon case.

Our results with the central values of the parameters are listed in Table~\ref{tab:xi1530}, which also shows the decay widths of $\Sigma(1385)$ 
and $\Delta(1232)$ resonances. 
Although our results show about 20\% discrepancy with the data, the dominant branching fractions could be reproduced. 
Considering the simplicity of the model, this is quite successful.
It should be noted that the decay width of $\Delta(1232)$ is significantly underestimated compared with the data.
Such a shortcoming is also observed in various versions of quark models~\cite{MLMP03,XZ13} and is ascribed to the significant role of meson clouds.

\begin{table}[t]
\centering
 \begin{ruledtabular}  \renewcommand{\arraystretch}{1.5}
\caption{\label{tab:xi1530} Predicted decay width of $\Xi(1530)$ of $J^P=3/2^+$ and other decuplet baryons in the units of MeV. 
We use the mass of initial and final states from PDG~\cite{PDG20}.
The branching fraction for each decay mode is indicated in parenthesis.  }
\begin{tabular}{cccccc}
	State & Channel & $\Gamma_{\rm NR}$  & $\Gamma_{\rm NR+RC}$ & $\Gamma_{\rm Expt.}$ \cite{PDG20}  \\ \hline
	$\Xi(1530)^0$	& $\Xi\pi$ & $19.9$ 	& $10.7$ & $9.1 \pm 0.5$\\   \hline 
	$\Sigma(1385)^+$ & $\Sigma \pi$ & $7.7$	 (13.4\%)	& $3.7$  (13.1\%) & $(11.7\pm 1.5 \%)$ \\
	                 & $\Lambda \pi$ & $49.7$	 (86.7\%)	& $24.6$  (86.9\%)	& $(87.0\pm 1.5 \%)$    \\
			& Sum & 57.3 & 28.3	& $36.0 \pm 0.7$ \\ \hline 
	$\Delta(1232)^+$ & $p\pi$	 & $124$  & $55.1$ & $\approx 117$ \\
\end{tabular}
\renewcommand{\arraystretch}{1}
\end{ruledtabular}
\end{table}

\begin{table*}[t]
\begin{ruledtabular}
\renewcommand{\arraystretch}{1.5}
\centering
\caption{\label{xi1690} 
Predicted decay width of $\Xi(1690)^0$ with various wave function assignments in the quark model in the units of MeV. 
We use the mass of initial and final states from PDG~\cite{PDG20}.
The branching fraction for each decay mode is given in parenthesis.}
\begin{tabular}{cccccccc}
$J^P$ &  State & Decay width & $\Xi\pi$ & $\Xi(1530)\pi$ & $\Lambda \bar{K}$ & $\Sigma \bar{K} $ & Sum   \\ \hline
\multirow{6}{*}{$\frac{1}{2}^-$	} &\multirow{2}{*}{$\ket{70,{}^{2}8,1,1,\tfrac{1}{2}^-}$} & $\Gamma_{\rm NR}$ & 2.0 (5.9\%)  
& 0.002 ($\approx 0$\%)  & 16.3 ($48.2\%$) & 15.5 (45.9\%)	 & 33.8    \\
 & & $\Gamma_{\rm NR+RC}$	  & 2.7 (7.6\%)	 & 0.001 ($\approx 0$\%)  & 17.2 (48.6\%)	  & 15.5 (43.8\%) & 35.4 \\ \cline{2-8} 
 &\multirow{2}{*}{$\ket{70,{}^{4}8,1,1,\tfrac{1}{2}^-}$}	& $\Gamma_{\rm NR}$   & 32.1 (61.4\%)	& 0.0005 ($\approx 0$\%)    & 16.3 (32.2\%)
 & 3.9 (7.4\%)	 &52.3	\\
 && $\Gamma_{\rm NR+RC}$	   &  42.5 (66.8\%)			  & 0.0003 ($\approx 0$\%)		&  17.2 (27.1\%)		 & 3.9 (6.1\%)				   
 & 63.6\\ \cline{2-8}
 &\multirow{2}{*}{$\ket{70,{}^{2}10,1,1,\tfrac{1}{2}^-}$}	  		& $\Gamma_{\rm NR}$  		& 2.0 (28.1\%)		 	 & 0.002 ($\approx 0$\%)	     
 & 4.1 (57.8\%)		 		   & 1.0 (14.1\%)      		   & 7.1 \\
& & $\Gamma_{\rm NR+RC}$	  & 2.7 (33.8\%)			& 0.001 ($\approx 0$\%)		 	& 4.3 (53.7\%)     		  	 & 1.0 (12.5\%)		  		 
& 8.0 \\ \hline
\multirow{6}{*}{$\frac{3}{2}^-$	} &\multirow{2}{*}{$\ket{70,{}^{2}8,1,1,\tfrac{3}{2}^-}$	}	 & $\Gamma_{\rm NR}$		    & 0.3 (20.0\%)			  	 
 & 1.0 (66.7\%)	  			& 0.2 (13.3\%)		 		 & $10^{-3}$ ($\approx 0$\%)	  & 1.5 \\
& & $\Gamma_{\rm NR+RC}$	   & 0.2 (14.3\%)		 	  	& 1.0 (71.4\%) 			   & 0.2 (14.3\%)				 
& $10^{-3}$ ($\approx 0$\%)	  & 1.4 \\ \cline{2-8}
&\multirow{2}{*}{$\ket{70,{}^{4}8,1,1,\tfrac{3}{2}^-}$}  			& $\Gamma_{\rm NR}$  		 & 0.5 (29.1\%)				 & 1.2 (69.8\%)				
& 0.02 (1.2\%)			& $10^{-5}$ ($\approx 0$\%) 	 &  1.7 \\
& & $\Gamma_{\rm NR+RC}$	   & 0.3 (19.8\%)				& 1.2 (78.9\%)			   & 0.02 (1.3\%) 				
& $10^{-5}$ ($\approx 0$\%)	& 1.5 \\ \cline{2-8}
&\multirow{2}{*}{$\ket{70,{}^{2}10,1,1,\tfrac{3}{2}^-}$}	       & $\Gamma_{\rm NR}$ 		   & 0.3 (22.0\%)				 	
& 1.0 (73.5\%)		  			& 0.06	(4.4\%)	 		& $10^{-4}$	($\approx 0$\%)	 & 1.4	\\
 &	& $\Gamma_{\rm NR+RC}$	  & 0.2	(14.7\%)				    & 1.0 (80.9\%)		   			& 0.05 (4.4\%) 		  		
 &  $10^{-4}$ ($\approx 0$\%)	 & 	1.3 \\ \hline 
\multirow{2}{*}{$\frac{5}{2}^-$	}&\multirow{2}{*}{$\ket{70,{}^{4}8,1,1,\tfrac{5}{2}^-}$}  	& $\Gamma_{\rm NR}$  		 & 2.9 (96.6\%)				  
& 0.002 (0.1\%)		   & 0.1 (3.3\%)		& $10^{-4}$ ($\approx 0$\%)	 & 3.0 \\
&	& $\Gamma_{\rm NR+RC}$	   & 1.9 (95.0\%)				 & 0.001 (0.1\%)		  &	0.1 (5.0\%)					
&  $10^{-4}$ ($\approx 0$\%)	& 2.0  \\ \hline
\multirow{2}{*}{$\frac{1}{2}^+$	} &\multirow{2}{*}{$\ket{ 56,{}^{2}8,2,0,\tfrac{1}{2}^+}$}  & $\Gamma_{\rm NR}$		 	& 0.2 (23.8\%)			 	  
& 0.02  (2.4\%)  	   & 0.4 (47.6\%)   			& 0.22	(26.2\%)			 & 0.8 \\
&	 & $\Gamma_{\rm NR+RC}$	    & 2.2 (50.0\%)			      & 0.2	 (4.6\%) 		   & 1.3 (29.5\%)		 		  
& 0.7	(15.9\%)	  		 	& 4.4 \\ \hline
&{\rm Expt.} \cite{PDG20}	 & 	&	&	&		&	& $<30$ \\
\end{tabular}
		\renewcommand{\arraystretch}{1}
\end{ruledtabular}
\end{table*}

\subsection{\boldmath ${\Xi(1690)}$ and ${\Xi(1620)}$ resonances}

The existence of low-lying $\Xi(1690)$ and $\Xi(1620)$ states is puzzling in the quark model because of their low masses,
and this observation induced various ideas~\cite{ROB02,GLN03,Oh07}.
The $\Xi(1690)$ resonance has a three-star rating in PDG with the decay width of $\Gamma<30$~MeV~\cite{PDG20}.  
In earlier experiments, the decay width of $\Xi(1690)^0$ is measured to be $44\pm23$~MeV or $20\pm4$~MeV~\cite{ACNO78}.
This uncertainty is removed by recent measurements when $\Gamma=27.1 \pm 10.0$~MeV and $\Gamma= 25.9 \pm 9.5$~MeV
were reported by the BESIII Collaboration~\cite{BESIII-15b} and the LHCb Collaboration~\cite{LHCb-20}, respectively. 
However its spin-parity is yet to be determined, although there is an evidence that it has $J^P=1/2^-$ from the interference pattern in the
analysis of the $\Lambda_c^+\to \Xi^+\pi^- K^+$ decay~\cite{BABAR08}.
Moreover, various models such as in the chiral-unitary model~\cite{ROB02,GLN03,KMHNNN16,Sekihara15}, 
Skyrme model~\cite{Oh07}, and QCD sum rules calculation~\cite{AAS18} favor $J^P=1/2^-$.
Therefore, further confirmation of its spin and parity is strongly called for to understand its structure and can be done by measuring 
polarization observables in the production through $K^-p$ scattering~\cite{NOH12}.

\begin{figure}[b]
	\centering
	\includegraphics[scale=0.25]{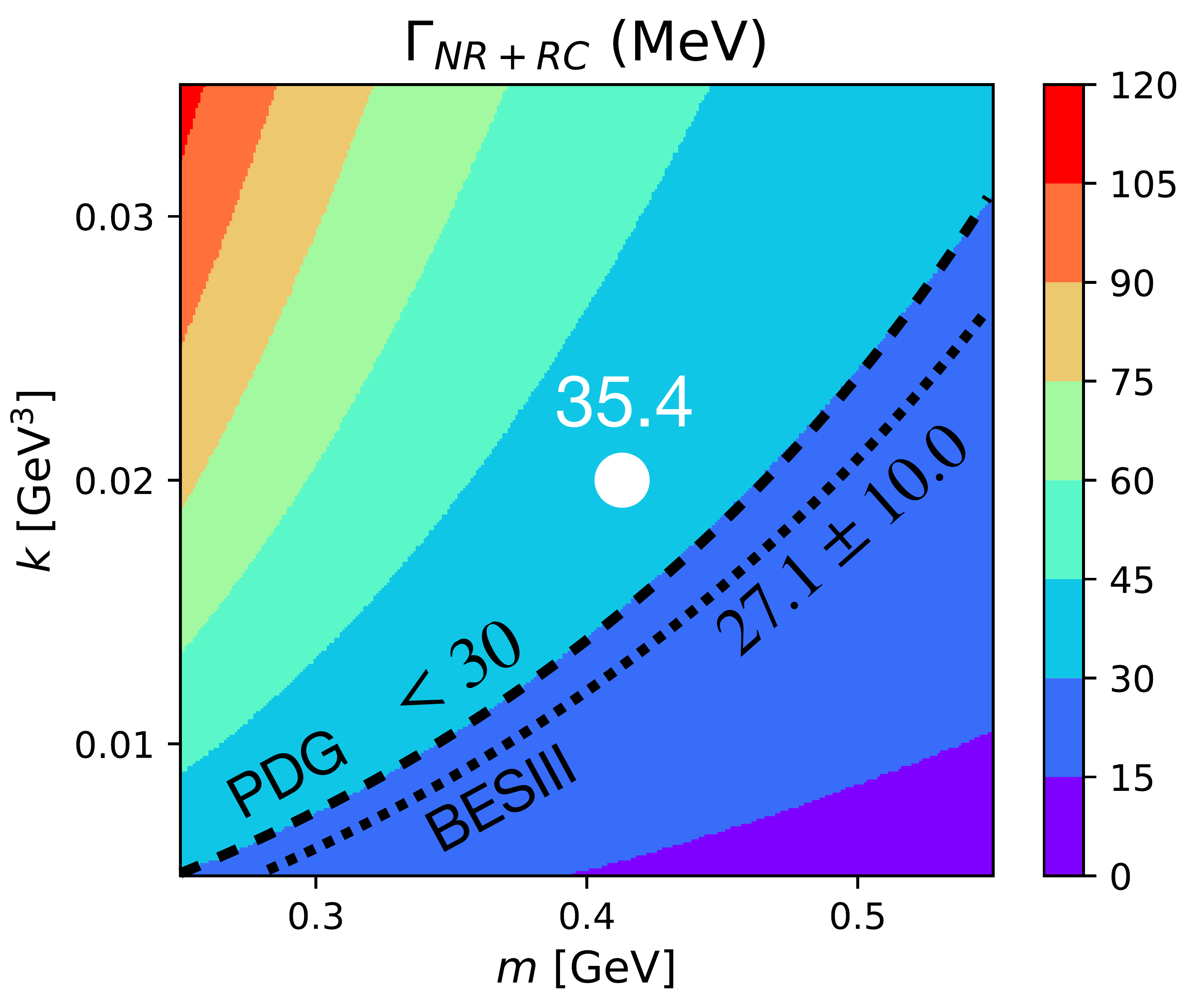}
	\caption{\label{xi1690fig} 
	Contour plot for the predicted decay width of $\Xi(1690)^0$ with the $\ket{70,^{2}8,1,1,\tfrac{1}{2}^-}$ assignment 
	as a function of the quark mass $m$ and spring constant $k$. 
	The black dashed line is the experimental data from PDG~\cite{PDG20}.}
\end{figure}

In the present study, we calculate the decay widths of this resonance in the quark model assuming the three-quark structure of the
resonance. 
In order to test the quantum numbers of $\Xi(1690)$ we consider various wave functions for this state and scrutinize its decay channels.
The obtained decay widths in the quark model with various spin and parity assignments are presented in Table~\ref{xi1690}.
Our results show that the relativistic corrections for the negative parity states are relatively small, which is consistent with our finding 
in heavy baryon cases~\cite{ASH21}.
Throughout this study, we found that the $\ket{70,{}^{2}8,1,1,\tfrac{1}{2}^-}$ configuration having $J^P=1/2^-$ may describe the 
decays of $\Xi(1690)$.
The predicted decay width of 35.4~MeV is compatible with the experimental data and its parameter dependence is shown 
in Fig.~\ref{xi1690fig} which shows the contour plot of the total decay width as a function of the quark mass and spring constant.

Another clue on the quantum numbers of $\Xi(1690)$ can be found in the branching fractions of its strong decays.
In Ref.~\cite{Belle-02}, the Belle Collaboration reported the branching ratio of the $\Xi(1690)^0$ decay rates as
 \begin{eqnarray}
	R^{\Sigma^+ K^-}_{\Lambda^0 \bar{K}^0}= \frac{\Gamma(\Sigma^+ K^-)}{\Gamma(\Lambda^0 \bar{K}^0)} = 0.50 \pm 0.26.
\end{eqnarray}
This is much smaller than the value of around 2.7 measured in 1970s~\cite{ACNO78}.
Our result, $R^{\Sigma \bar{K} }_{\Lambda \bar{K} }=0.9$ is in fair agreement with the value reported by the Belle Collaboration.
Moreover, the obtained ratios are consistent with the observed values
 \begin{eqnarray}
	R^{\Xi \pi}_{\Sigma \bar{K} } < 0.09,  \qquad
	R^{\Xi(1530)\pi}_{\Sigma \bar{K} } < 0.06
\end{eqnarray}
of Ref.~\cite{ACNO78}.
As shown in Table~\ref{xi1690}, only the configuration of $\ket{70,{}^{2}8,1,1,\tfrac{1}{2}^-}$ has a large branching fraction to the $\Sigma \bar{K}$ 
channel over the $\Xi\pi$ and $\Xi(1530)\pi$ channels.
In Ref.~\cite{XZ13}, the mixing scenario was investigated to get a better fit with the branching fractions.
It is found that the dominant contribution is from the  $\ket{70,{}^{2}8,1,1,\tfrac{1}{2}^-}$ configuration which is consistent with the present study.

It is worth noting that both the $\ket{ 70,{}^{2}8,1,1,\tfrac{1}{2}^-}$ and $\ket{70,{}^{2}10,1,1,\tfrac{1}{2}^-}$ states predict the same partial 
decay widths for $\Xi\pi$ and $\Xi(1530)\pi$ channels because of the identical spin-flavor factor.
However, they have very different partial decay widths for the $\Lambda \bar{K} $ and $\Sigma \bar{K} $ channels.
Furthermore, the $\ket{ 70,{}^{4}8,1,1,\tfrac{1}{2}^-}$ state decays dominantly into the $\Xi\pi$ channel.
Thus, only the $\ket{70,^{2}8,1,1,\tfrac{1}{2}^-}$ configuration can reproduce not only the decay width but the observed ratios as well.
We found that the other spin-parity assignments including the radial excitation cannot reproduce the observed decay patterns.

We also consider the decays of the $\Xi(1620)$ resonance that is recently confirmed in the $\Xi_c^+ \rightarrow \Xi^-\pi^+\pi^+$ decay 
by the Belle Collaboration~\cite{Belle-18b}.
It still has a one-star rating with a decay width of around 20-40 MeV~\cite{PDG20}.
This means that it has a larger decay width despite having a smaller phase space factor than the case of $\Xi(1690)$.
Therefore, it is anticipated from Table~\ref{xi1690} that the quark configuration of $\ket{70,^{4}8,1,1,\tfrac{1}{2}^-}$ can give a larger
decay widths and it would be natural to assign this configuration to $\Xi(1620)$.
Other configurations give very small decay width contrary to the observation.
The computed decay widths of $\Xi(1620)$ within this configuration are presented in Table~\ref{xi1620}.
It should also be noted that the $J^P=1/2^-$ assignment is consistent with the predictions of other models such as the 
Skyrme model~\cite{Oh07} or chiral-unitary model~\cite{ROB02,GLN03} in spite of having different structures for these states.

\begin{table}[t]
\begin{ruledtabular}
\renewcommand{\arraystretch}{1.5}    \centering
\caption{\label{xi1620} 
Predicted decay widths of $\Xi(1620)^0$ with $J^P=1/2^-$ and $\ket{70,{}^{4}8,1,1,\tfrac{1}{2}^-}$ assignment in the units of MeV.
We use the mass of initial and final states from PDG~\cite{PDG20}.
The branching fraction for each decay mode is indicated in parentheses.}
\begin{tabular}{cccccc}
State & 	Channel	 &  $\Gamma_{\rm NR}$ & $\Gamma_{\rm NR+RC}$ & 	$\Gamma_{\rm Expt.}$ \cite{PDG20} \\ \hline
$\Xi(1620)^0$	  & 	$\Xi\pi$ &	23.8 (80.9\%)	 & 29.0 (83.8\%)	& 	 \\ 
& 	$\Lambda \bar{K} $	&	5.6 (19.1\%) &	5.6	 (16.2\%)	& 		 \\ 
& 	Sum	& 29.4	& 34.6 & 	$40\pm15$	 \\
\end{tabular}
\renewcommand{\arraystretch}{1}
\end{ruledtabular}
\end{table}

\subsection{\boldmath ${\Xi(1820)}$ and Roper-like  ${\Xi}$ resonance}

Among the observed $\Xi$ resonances, the $\Xi(1820)$ is the only one whose quantum numbers are determined experimentally.
Its mass is about 500~MeV larger than the ground state so it lies in the region of the anticipated Roper state analog.
(See Fig.~\ref{roper}.)
However, it is found to have $J^P = 3/2^-$~\cite{BBBB87b,TDDG78} so that it is excluded from a candidate for the Roper-like state.
According to PDG, this resonance has the estimated decay width of $\Gamma= 24^{+15}_{-10}$ MeV~\cite{PDG20}.
However, the data span a wide range.
For example, the old data of 1970s and 1980s cover from about 12~MeV to about 103~MeV.
Therefore, new measurements at modern facilities have been strongly called for.

Recently, the LHCb Collaboration found a rather large width of $36.0 \pm 4.0$~MeV in the decay of 
$\Xi_b\to J/\varPsi \Lambda K$~\cite{LHCb-20}.
Meanwhile, the BESIII Collaboration observed an even larger width of $54.4 \pm15.7$~MeV in the analysis of 
$\varPsi(3686)\to K^-\Lambda\bar{\Xi}^+$~\cite{BESIII-15b}.
However, they also measured a rather smaller width of $17.0 \pm 15.0$~MeV in the process of 
$e^+e^-\to \Xi^-\bar{\Xi}^+$~\cite{BESIII-19}. 
Therefore, although the BESIII data have large experimental error bars, the issue of scattered data points is still not resolved and
more precise measurements in various processes are required.
On the other hand, the wide range of the measured decay widths may suggest the possibility of the overlap with other resonances
in this mass region.
Although its spin and parity was found to be $3/2^-$~\cite{BBBB87b,TDDG78}, it cannot be excluded that the data might be 
contaminated by other nearby resonances such as the Roper-like state with $J^P=1/2^+$ or yet unidentified resonances.

We compute the decay widths of $\Xi(1820)$ with various configurations and the results are tabulated in Table~\ref{xi1820}.
We first check the role of the relativistic corrections.
Similar to the results of the previous subsection, we observe that the relativistic corrections are relatively small (less than 20\%) 
for the negative parity states.
However, with the same phase space factor, the decay width for the $J^P=1/2^+$ assignment experiences a large increase.
Again, this is consistent with what we have found in the heavy baryon case~\cite{ASH21}.

\begin{table*}[t]
\begin{ruledtabular}
\renewcommand{\arraystretch}{1.5}
\centering
\caption{\label{xi1820} 
Predicted decay width and branching fraction of $\Xi(1820)^0$ with various quark model assignments in the units of MeV. 
We use the mass of initial and final states from PDG~\cite{PDG20}.
The branching fraction for each decay mode is indicated in parentheses.}
\begin{tabular}{cccccccc}
 &	State &  &$\Xi\pi$	   & $\Xi(1530)\pi$	& $\Lambda \bar{K} $ & $\Sigma \bar{K} $  & Sum	\\ \hline 
\multirow{6}{*}{$\frac{1}{2}^-$	}	&  \multirow{2}{*}{$\ket{70,{}^{2}8,1,1,\tfrac{1}{2}^-}$	}	 & $\Gamma_{\rm NR}$  		  & 2.5	(2.6\%)  		& 0.7  (0.7\%)		& 18.3  (18.9\%)	 	 & 75.5	(77.8\%)		 & 97.0	\\
			&																						  & $\Gamma_{\rm NR+RC}$	  & 3.9	 (3.6\%)		  & 0.5   (0.4\%)	 	      & 21.5  (19.6\%)				  & 83.7 (76.4\%)			 & 110 \\ \cline{2-8}
			&\multirow{2}{*}{$\ket{70,{}^{4}8,1,1,\tfrac{1}{2}^-}$}			 & $\Gamma_{\rm NR}$  		  & 39.3 (51.3\%)		& 0.2 (0.3\%)		   & 18.3 (23.8\%)			& 18.8	(24.5\%)		  & 76.6 	\\
			&																						 & $\Gamma_{\rm NR+RC}$	  & 62.4 (59.4\%)		 & 0.1	(0.1\%)		 & 21.5	(20.5\%)			 & 20.8	(19.8\%)		   & 105	 \\ \cline{2-8}
			&\multirow{2}{*}{$\ket{70,{}^{2}10,1,1,\tfrac{1}{2}^-}$}		 	& $\Gamma_{\rm NR}$ 		& 2.5 (20.2\%)	 		& 0.7 (5.6\%)		 	& 4.6 (37.1\%)	  			& 4.7 (37.9\%)    		   & 12.5 	  \\
			&																						 & $\Gamma_{\rm NR+RC}$	  & 3.9	(26.2\%)		   & 0.5 (3.4\%)		  & 5.4 (36.2\%)		 	  & 5.2 (34.9\%)	 		 & 15.0 	\\ \hline 
			\multirow{6}{*}{$\frac{3}{2}^-$	}&\multirow{2}{*}{$\ket{70,{}^{2}8,1,1,\tfrac{3}{2}^-}$	} & $\Gamma_{\rm NR}$ 		     & 1.2	(8.5\%)  		  & 6.2  (44.0\%)		& 2.6 (18.4\%)	 & 4.1 	 (29.1\%)  		   & 14.1	  \\
			&																						  & $\Gamma_{\rm NR+RC}$   & 0.8 (5.6\%)	   		& 7.2 	(50.7\%)	  & 2.4	(16.9\%)			   & 3.8 (26.7\%)	  		 & 14.2 	\\ \cline{2-8}
			&\multirow{2}{*}{	$\ket{70,{}^{4}8,1,1,\tfrac{3}{2}^-}$	}		 & $\Gamma_{\rm NR}$  	     &  1.9	(26.8\%)		  & 4.8 (67.6\%)		& 0.3 (4.2\%)				 & 0.1	(1.4\%)			& 7.1		\\
			&																						  & $\Gamma_{\rm NR+RC}$	   &  1.3 (16.0\%)	 		& 6.5 (80.2\%)		  & 0.2	(2.5\%)				&	0.1 (1.2\%)			  &	8.1		\\ \cline{2-8}			
			&\multirow{2}{*}{$\ket{ 70,{}^{2}10,1,1,\tfrac{3}{2}^-}$	}	 	& $\Gamma_{\rm NR}$		    & 1.2 (14.6\%)	 		 & 6.2 (75.6\%)			& 0.7 (8.5\%) 				 & 0.2 (2.4\%)		 	   & 8.3	\\
			&																						 & $\Gamma_{\rm NR+RC}$   	& 0.8 (9.1\%)	  		& 7.2 (81.8\%)			& 0.6 (6.8\%)		 		& 0.2 (2.3\%)	 		 & 8.8		\\ \hline 
			\multirow{2}{*}{$\frac{5}{2}^-$	}&\multirow{2}{*}{	$\ket{70,{}^{4}8,1,1,\tfrac{5}{2}^-}$	}		& $\Gamma_{\rm NR}$    & 11.3 (79.6\%)		 & 0.7 	(4.9\%)		 &	1.6 (11.3\%)				  &  0.6 (4.2\%)		   & 14.2	 \\
			&																						  & $\Gamma_{\rm NR+RC}$	 & 7.6	(76.0\%)		& 0.5 (5.0\%)			& 1.5 (15.0\%)				 &0.5 (5.0\%)			  & 10.1	 \\ \hline 
			\multirow{2}{*}{$\frac{1}{2}^+$	} &\multirow{2}{*}{	$\ket{56,{}^{2}8,2,0,\tfrac{1}{2}^+}$	}	    & $\Gamma_{\rm NR}$			  & 0.3	(1.5\%)  		& 0.9 (4.5\%)	 		& 0.9 (4.5\%)	& 17.8	 (89.5\%)		& 19.9		\\
			&																						  & $\Gamma_{\rm NR+RC}$	  & 4.5 (5.4\%)  		& 8.8 (10.5\%) 			& 3.9 (4.6\%)			 	&  66.8	(79.5\%)	 	& 84.0 		\\ \hline
			&	{\rm Expt.} \cite{PDG20,ABFM69} 						& 											 &		$(10 \pm 10\%)$			  &	$(30 \pm 15\%)$		&	$(30 \pm 15\%)$	 &	$(30 \pm 15\%)$				& $24^{+15}_{-10}$ \\
		\end{tabular}
		\renewcommand{\arraystretch}{1}
	\end{ruledtabular}
\end{table*}

As shown in Table~\ref{xi1820}, all the states of $J^P = 3/2^-$ and $5/2^-$ and one of $J^P = 1/2^-$ 
(the $\ket{70,{}^{2}10}$ configuration) have small decay widths of around 10~MeV.
Instead, large decay widths of around 100~MeV are predicted for the
$J^P=1/2^+$ state and for $\ket{70,{}^{2}8}$ and $\ket{70,{}^{4}8}$ of $J^P = 1/2^-$.
As for the branching fractions, the $1/2^-$ states generally have a small fraction to the $\Xi(1530)\pi$ channel as a result of the $d$-wave nature.
On the other hand, the $3/2^-$ states decay into $\Xi(1530)\pi$ in $s$-wave making it a dominant decaying channel.
However, the $5/2^-$ and $1/2^+$ states decay to all decay modes in $d$-wave and $p$-wave, respectively.
For the $5/2^-$ state, a large branching fraction to the $\Xi\pi$ channel is observed, while the decay of the $1/2^+$ state is dominated by 
the $\Sigma \bar{K}$ channel.

\begin{figure}[t]
	\centering
	\includegraphics[scale=0.25]{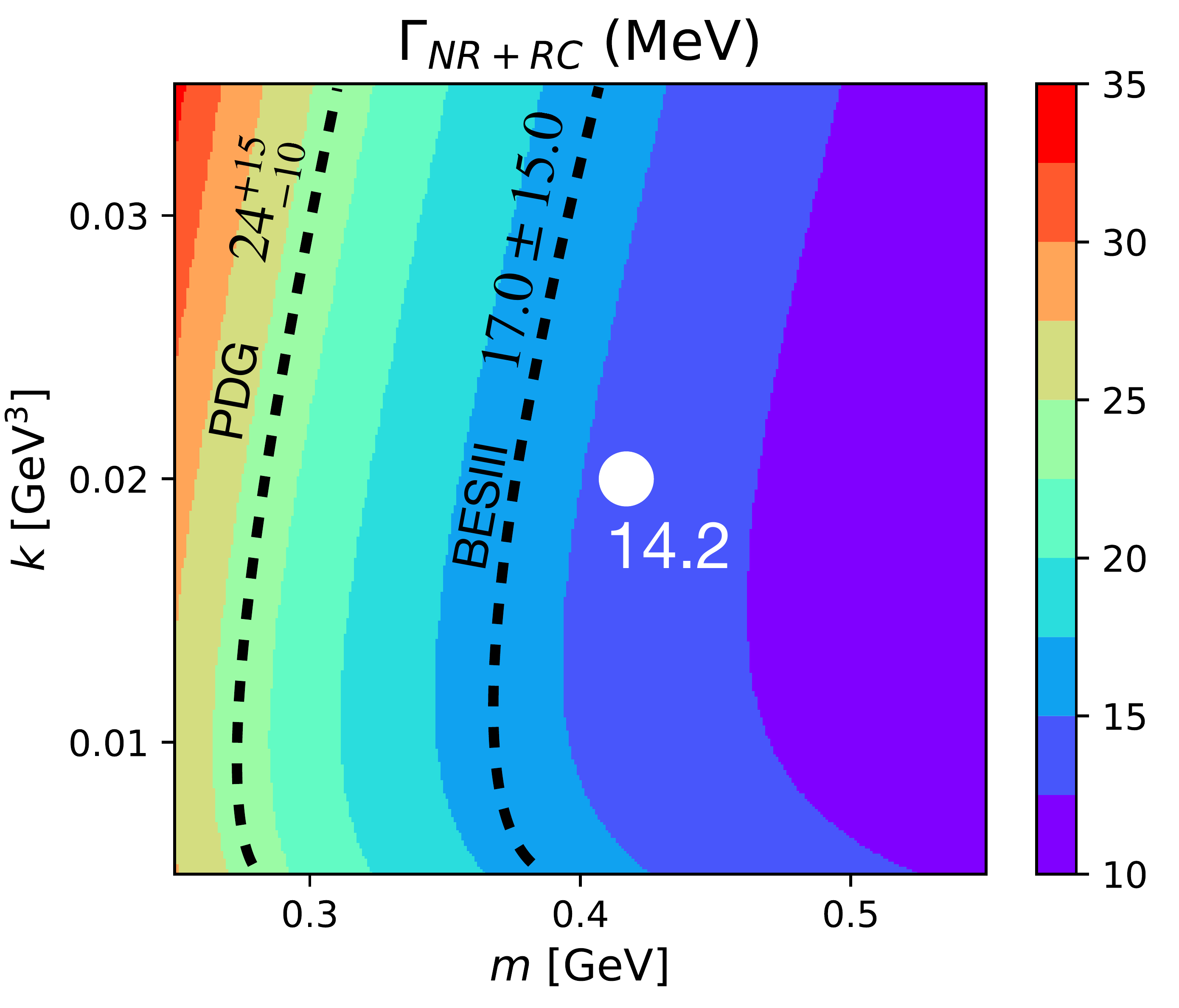}
	\caption{\label{xi1820fig} 
	Contour plot for the decay width of $\Xi(1820)^0$ with the $\ket{70,{}^{2}8,1,1,\tfrac{3}{2}^-}$ configuration 
	as a function of the quark mass $m$ and spring constant $k$. 
	The black dashed lines are the experimental data from PDG~\cite{PDG20} and BESIII~\cite{BESIII-19}.}
\end{figure}

For further exploration, we consider the branching ratios of the $\Xi(1820)$ decays. 
The branching ratios of $\Xi(1820)$ decay was measured in the $K^- p$ scatterings in 1960s~\cite{ABFM69}.
The measured data read
\begin{subequations}
\begin{eqnarray}
\mathcal{B}(\Xi\pi) &=& 0.1 \pm 0.1, \\
\mathcal{B}(\Xi(1530)\pi) &=& 0.30 \pm 0.15, \\
\mathcal{B}(\Lambda \bar{K}) &=& 0.30 \pm 0.15, \\
\mathcal{B}(\Sigma \bar{K})  &=& 0.30 \pm 0.15.
\end{eqnarray}
\end{subequations}
From Table~\ref{xi1820}, we find that the $\ket{70,{}^{2}8,1,1,\tfrac{3}{2}^-}$ configuration has a fair agreement with the data.
In order to have a better fitting, mixing with other $3/2^-$ configurations would be considered, but the study of Ref.~\cite{XZ13} 
indicates that the $\ket{70,{}^{2}8,1,1,\tfrac{3}{2}^-}$ configuration is the dominant.
We also note that the measured branching fractions are not compatible with the $5/2^-$ or $1/2^-$ states.
In particular, the $5/2^-$ state has a large branching ratio of $\mathcal{B}(\Xi\pi)=76\%$ while $1/2^-$ state has a small value
of $\mathcal{B}(\Xi(1530)\pi)=3.4\%$.
Such values are not consistent with the data and we conclude that the observed $\Xi(1820)$ with a relatively small decay width 
is compatible with $J^P=3/2^-$.
Figure~\ref{xi1820fig} shows the contour plot of the decay width as a function of model parameters.

\begin{figure}[t]
	\centering
	\includegraphics[scale=0.25]{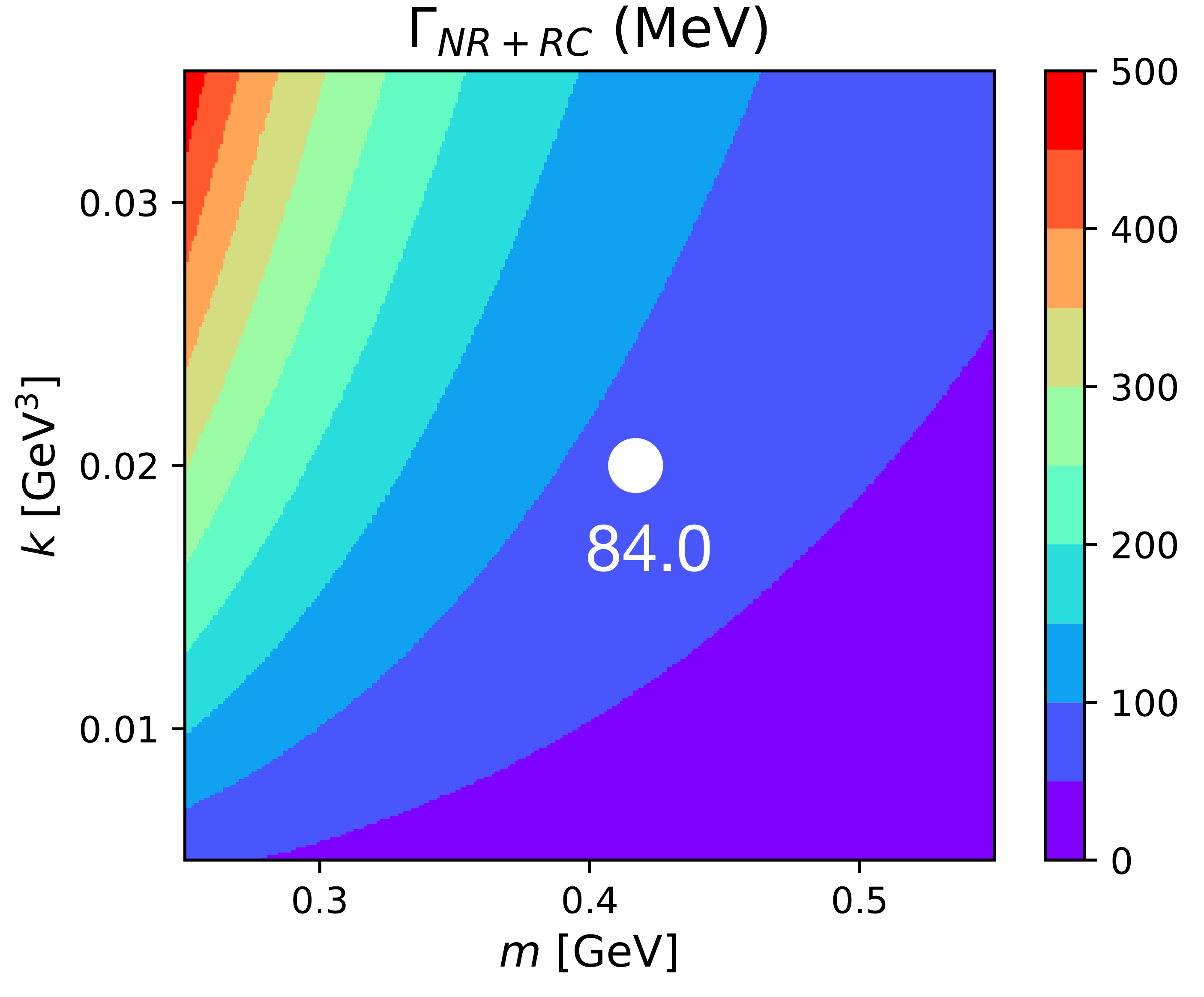}
	\caption{\label{xi1840fig} 
	Contour plot of the predicted decay width of the Roper-like state $\ket{56,^{2}8,2,0,\tfrac{1}{2}^+}$ as a function of 
	the quark mass $m$ and spring constant $k$. 
	The mass of this state is assumed to be 1823~MeV.}
\end{figure}

Being motivated by the dispersion of the data, we also consider the scenario that other resonances exist in the mass region of 1.8~GeV.
Table~\ref{xi1820} shows that the $\ket{70,{}^{2}8}$ and $\ket{70,{}^{4}8}$ states of $J^P = \frac12^-$ may have large decay widths.
However, these states are already identified as candidates for $\Xi(1690)$ and $\Xi(1620)$.
Then the other candidate which may have a large decay width is the Roper-like state with $J^P=1/2^+$.
In the present model study, when we assume that the mass of this state is 1823~MeV, its decay width is predicted to be about 80~MeV,
and it becomes even larger if we assume a higher mass.
The obtained decay width of this state as a function of the quark model parameters $m$ and $k$ is given as a contour plot 
in Fig.~\ref{xi1840fig}, which shows that the decay width is sensitive to model parameters.
Furthermore, the role of the relativistic corrections is nontrivial. 
For comparison with the cases of other Roper and Roper-like states with strangeness $S=0$ or $S=-1$, we present the decay
widths of $N(1440)$, $\Lambda(1600)$, and $\Sigma(1660)$ in Table~\ref{roper-like}.

Although we have assigned $J^P=1/2^-$ states to $\Xi(1690)$ and $\Xi(1620)$, it may be plausible that the large decay width observed
in a few old experiments may indicate another resonance of $J^P = 1/2^-$.
In order to distinguish these scenarios, we consider the ratios of branching fractions.
Table~\ref{xi1820} shows that both the $\ket{70,{}^{2}8}$ of $J^P = 1/2^-$ and the radially excited $J^P = 1/2^+$ states have the dominant
decay channel of $\Sigma\bar{K}$.
Therefore, if a resonance is discovered around 1.8~GeV in the $\Sigma \bar{K}$ channel, its quantum number cannot be easily identified.
However, we find that the ratio of branching fractions, 
\begin{eqnarray}
	R^{\Xi\pi}_{\Xi(1530)\pi} = \frac{\Gamma \bm{(} \Xi(1820) \to \Xi\pi \bm{)} }{\Gamma \bm{(} \Xi(1820) \to \Xi(1530)\pi \bm{)} }	
\end{eqnarray} 
may be useful to distinguish the spin-parity quantum numbers as in the heavy baryon sector~\cite{ANHT20a}.
Table~\ref{xi1820} indicates that all the $1/2^-$ states have very suppressed decay into $\Xi(1530) \pi$ and the ratio has a quite large value,
while for the $1/2^+$ state, it is the opposite and the ratio is less than 1.%
\footnote{Our estimates from Table~\ref{xi1820} are $R^{\Xi\pi}_{\Xi(1530)\pi} \approx 7.8$, $620$, $7.8$ for $J^P = 1/2^-$ states and
0.5 for $J^P = 1/2^+$ state.}
Therefore, a large width with dominant $\Sigma \bar{K}$ and $R^{\Xi\pi}_{\Xi(1530)\pi}$ close to unity would be good signatures 
of the Roper-like $\Xi$ resonance.

\begin{table}[t]
	\begin{ruledtabular}
		\renewcommand{\arraystretch}{1.2}
		\centering
		\caption{\label{roper-like} Predicted decay widths of the Roper resonance and its siblings in the quark model in the units of MeV. 
		We use the mass of initial and final states from PDG~\cite{PDG20}.
		The branching fraction for each decay mode is indicated in parentheses.}
		\begin{tabular}{ccccccc}
			\multirow{2}{*}{State }	 &  \multirow{2}{*}{Channel} 			  &\multirow{2}{*}{$\Gamma_{\rm NR}$} & \multirow{2}{*}{$\Gamma_{\rm NR+RC}$} &  \multicolumn{2}{c}{$\Gamma_{\rm Expt.}$~\cite{PDG20}} &	\\
			&								&							&				&			(Pole) 			&  (BW)  \\ \hline
			$N(1440)$		 		 & $N\pi$					&  12.1	(91.7\%)  & 151 (92.1\%) 	  & \multicolumn{2}{c}{$(55 - 75\%)$}\\
			& $ \Delta\pi$			  &	  1.1 (8.3\%) 	& 12.9	(7.9\%)  & \multicolumn{2}{c}{$(6 - 27\%)$}\\
			& Sum						&	13.2  	 & 164	  & $\approx 175$ & $\approx 350$ \\ \hline 
			$\Lambda(1600)$		&  	$\Sigma\pi$		 	&   4.0	(19.9\%) 	& 41.3	(37.0\%) & \multicolumn{2}{c}{$(10 - 60\%)$}\\		
			&  $\Sigma(1385)\pi$		& 	0.8	(4.0\%) 				    & 9.4  (8.5\%) 	  &  \multicolumn{2}{c}{$(9 \pm 4\%)$}\\
			&   $ N\bar{K}$    		 & 	15.3 (76.1\%) 					& 60.9	(54.5\%)   & \multicolumn{2}{c}{$(15 - 30\%)$}\\
			& Sum						&	20.1				   & 112	   &$\approx 180$  & $\approx 200$\\ \hline 
			$\Sigma(1660)$		  &  $\Sigma\pi$		   &  4.3	(59.7\%)   & 50.7	(60.5\%)  & \multicolumn{2}{c}{$(37 \pm 10\%)$}\\
			&  $\Sigma(1385)\pi$		& 	0.6	(8.3\%) 	 & 6.1	(7.3\%) 	   & \multicolumn{2}{c}{-}\\
			&  $ \Lambda\pi$ 		& 1.6	(22.2\%) 	  &	24.0 (28.6\%) 	  & \multicolumn{2}{c}{$(35 \pm 12\%)$}\\
			& $ N\bar{K}$ 			 & 0.7 (9.2\%) 	& 3.0 (3.6\%) 				&	\multicolumn{2}{c}{$(5 - 15\%)$}\\
			& 	Sum						 & 7.2		 & 83.8	   &  $290^{+140}_{-40}$ &  $\approx 200$ \\ 
		\end{tabular}
		\renewcommand{\arraystretch}{1}
	\end{ruledtabular}
\end{table}

Before closing this subsection, we mention some similarities between $\Xi$ and $\Xi_c$ baryon spectra. 
In the spectrum of $\Xi_c$ baryons, there are two nearby states: $\Xi_c(2970)$ which is claimed to be the Roper-like 
state~\cite{Belle-20b} and $\Xi_c(2965)$ which is recently observed by the LHCb Collaboration~\cite{LHCb-20b}.
These states were discovered in different channels, namely, the $\Xi_c(2970)$ was found in the $\Xi_c \pi\pi$ invariant mass
while the $\Xi_c(2965)$ was in the $\Lambda_c \bar{K} $ invariant mass.
In the heavy baryon sector, the Roper-like $\Xi_c$ resonance cannot decay into the $\Lambda_c \bar{K} $ channel
because of the spin selection rule~\cite{ANHT20b}, which excludes the possibility that the $\Xi_c(2965)$ 
could be a Roper-like state.
Furthermore, it was reported that $\Xi_c(2970)$ and $\Xi_c(2965)$ have different decay widths, namely,
$\Gamma[\Xi_c(2970)^0]=28^{+3.4}_{-4.0}$~MeV and $\Gamma[\Xi_c(2965)^0]=14.1\pm0.9$~MeV.
In the quark model, $\Xi_c(2965)$ with a smaller width can be accommodated as a $P$-wave excitation or the negative parity 
state~\cite{WXZ20}.
In the case of strangeness $S=-2$ resonance, however, there is no such spin selection rule and the $\bar{K} \Lambda$ decay mode is 
allowed as shown in Table~\ref{xi1820}.
However, as the case of $\Xi_c$ baryon spectrum indicates, one cannot exclude the scenario 
that the Roper-like $\Xi$ resonance would exist in the mass region of 1.8~GeV.

\begin{table*}[t]
\begin{ruledtabular}
\renewcommand{\arraystretch}{1.5}
\centering
\caption{\label{xi1950} 
Predicted decay width of $\Xi(1950)^0$ with various quark model configuration in the units of MeV. 
We use the mass of initial and final states from PDG~\cite{PDG20}.
The branching fraction for each decay mode is indicated in parenthesis. }
\begin{tabular}{ccccccccc}
			&State 				  																  &  			   							&$\Xi\pi$	 	   & $\Xi(1530)\pi$	 & $\Lambda \bar{K} $ & $\Sigma \bar{K} $   & $\Sigma(1385) \bar{K} $   & Sum\\ \hline
			\multirow{2}{*}{$\frac{1}{2}^-$	} &\multirow{2}{*}{$\ket{70,{}^{2}10,1,1,\tfrac{1}{2}^-}$	}	& $\Gamma_{\rm NR}$ 		& 2.2 (12.6\%)  & 4.6 (26.4\%) 	    & 3.5 (20.1\%)		 & 4.3 (24.8\%)	 		& 2.8 (16.1\%)	  & 17.4 \\
			&																					 	 & $\Gamma_{\rm NR+RC}$	  & 4.3 (21.5\%)		 	   & 3.0 (15.0\%) 	  & 4.7 (23.5\%)	   			 & 5.4 (27.0\%)   			& 2.6 (13.0\%)	& 20.0 \\ \hline
			\multirow{4}{*}{$\frac{3}{2}^-$	} &\multirow{2}{*}{$\ket{70,{}^{4}8,1,1,\tfrac{3}{2}^-}$}  			& $\Gamma_{\rm NR}$  		 & 4.5 (12.5\%)		 & 4.8 (13.3\%)			& 0.8 (2.2\%)		& 0.5 (1.4\%) 	 & 25.5 (70.6\%)	&  36.1 \\
			&																						 & $\Gamma_{\rm NR+RC}$	   & 3.0 (7.4\%)				& 9.3 (22.9\%)			   & 0.7 (1.7\%) 		& 0.4 (1.0\%) & 27.2 (67.0\%)	& 40.6\\ \cline{2-9}
			&\multirow{2}{*}{$\ket{70,{}^{2}10,1,1,\tfrac{3}{2}^-}$}	       & $\Gamma_{\rm NR}$ 		   & 2.8 (2.4\%)				 	& 11.7 (9.9\%)		  			& 1.9	(1.6\%)	 		& 1.2 (1.0\%)&	99.9 (84.7\%)	 & 118	\\
			&																						& $\Gamma_{\rm NR+RC}$	  & 1.9	(1.5\%)				    & 14.9 (12.0\%)		   			& 1.8 (1.5\%) 		   &  1.1 (0.9\%)& 104 (83.9\%)		 & 	124\\ \hline 
			\multirow{2}{*}{$\frac{5}{2}^-$	}&\multirow{2}{*}{$\ket{70,{}^{4}8,1,1,\tfrac{5}{2}^-}$}  	& $\Gamma_{\rm NR}$  		 & 26.7 (67.4\%)		 & 4.8 (12.1\%)		   & 4.6 (11.6\%)		& 2.8 (7.1\%)	 & 0.7 (1.8\%)	& 39.6 \\
			&																						& $\Gamma_{\rm NR+RC}$	   &18.1 (62.6\%)				 & 3.2 (11.1\%)		  &	4.3 (14.9\%)					&  2.6 (9.0\%)	& 0.7 (2.4\%)	& 28.9 \\ \hline
			\multirow{2}{*}{$\frac{1}{2}^+$	} &\multirow{2}{*}{$\ket{ 56,{}^{2}8,2,0,\tfrac{1}{2}^+}$}  & $\Gamma_{\rm NR}$		 	& 0.2 (0.7\%)			 	  & 2.5  (8.6\%)  	   & 0.7 (2.4\%)   & 22.6	(77.7\%)		&	3.1 (10.6\%)	 & 29.1 \\
			&																					 	 & $\Gamma_{\rm NR+RC}$	    & 6.0 (3.5\%)			      & 25.9 (15.2\%) 		   & 4.9 (2.9\%)		 		  & 118	(69.5\%)	  &		15.1 (8.9\%)	 	& 170 \\ \hline
			\multirow{2}{*}{$\frac{3}{2}^+$	} &\multirow{2}{*}{$\ket{ 56,{}^{2}10,2,0,\tfrac{3}{2}^+}$}  & $\Gamma_{\rm NR}$		 	& 0.8 (6.8\%)			 	  & 1.5  (12.8\%)  	   & 2.8 (23.9\%)   			& 1.5 (12.8\%)	&	5.1 (43.6\%)		 & 11.7 \\
			&																					 	 & $\Gamma_{\rm NR+RC}$	    & 24.0 (22.6\%)			      & 16.2	 (15.3\%) 		   & 19.6 (18.5\%)		 		  & 7.9	(7.5\%)	  	&	37.8 (35.7\%)	 	& 106\\ \hline
			& Expt. \cite{PDG20}						 & 					&						&					&						&						& & $60\pm 20$ \\
\end{tabular}
\renewcommand{\arraystretch}{1}
\end{ruledtabular}
\end{table*}

\subsection{\boldmath ${\Xi(1950)}$ resonance}

In PDG, $\Xi(1950)$ is rated as a three-star resonance with $\Gamma=60\pm 15$ MeV, but its spin and parity 
are unknown~\cite{PDG20}.
As mentioned by PDG, $\Xi$ resonances observed in the mass region
between 1875 and 2000 MeV are collected as $\Xi(1950)$ and it would be possible
that there are more than one $\Xi$ resonance near this mass as predicted by quark models.%
\footnote{In Ref.~\cite{PXN11}, it was suggested that there are three $\Xi$ resonances in this mass region which have $J^P = 1/2^-$,
$5/2^+$, and $5/2^-$, respectively. See also Refs.~\cite{XZ13,GP05b}.}
In fact, as shown in Table~\ref{total}, there are seven negative parity states but only three of them are observed as we have discussed
in the previous subsections.
In addition, one cannot exclude the case that positive parity states may also exist in this mass region.
In the present work, we explore the possibility that the radially excited state of the decuplet $\Xi(1530)$ with $J^P=3/2^+$ has a mass 
around 1.95~GeV being motivated that the Roper-like state would have a mass about 400-500~MeV higher than the corresponding ground
state.

The calculated decay width of $\Xi(1950)$ with various quark configurations are presented in Table~\ref{xi1950}.
The general behaviors of the predicted widths are similar to what we have discussed for $\Xi(1820)$ in the previous subsection, 
except that the $\Sigma(1385)\bar{K}$ channel is now open. 
It is found that the $J^P = 3/2^-$ states have the dominant $\Sigma(1385) \bar{K}$ channel partly due to the $s$-wave nature.
The $\ket{70,{}^{2}10,1,1,\tfrac{3}{2}^-}$ configuration thus becomes very broad.
We also find that the $5/2^-$ state has a dominant $\Xi\pi$ channel.
This observation would be helpful in searching for negative parity resonances in this mass region.

Since we are also interested in the Roper-like states, we consider the radial excitations of $\Xi$ baryons.
There can be two radially excited states of $\Xi$ with $1/2^+$ and $3/2^+$ that belong to the octet and decuplet baryons, respectively, 
as listed in Table~\ref{total}.
If we assume that the first radially excited $J^P = 1/2^+$ state has a mass around 1.95~GeV instead of 1.8~GeV, we would have a decay
width of 170~MeV, which is much larger than the estimated width of about 60~MeV~\cite{PDG20} on top of that its mass is higher than
the ground state by 640~MeV.
Therefore, it would be difficult to interpret $\Xi(1950)$ as the first radially excited state with $J^P = 1/2^+$.
Instead, we consider the first radially excited state of $J^P = 3/2^+$. 
Then the mass difference between $\Xi(1950)$ and $\Xi(1530)$ is about 420~MeV and the decay width is estimated to be about 100~MeV.
This state also decays to each mode in $p$ wave, but the branching fraction to the $\Sigma \bar{K}$ channel is observed to be small.
Since there may exist several $\Xi$ resonances, including negative and positive parity states, more precise measurements are required 
to clarify the issue.

\section{ $\mathbf{\Omega}$ baryon decays} \label{sec:Omega}

In this section, we discuss the low-lying $\Omega$ resonances in the quark model including the newly observed $\Omega(2012)$ 
and its LS partner.
We also discuss its first radial excitation that could be discovered in future experiments~\cite{KLF-20,JPARC-21}.

\subsection{\boldmath ${\Omega(2012)}$ resonance}

The $\Omega(2012)$ resonance was observed by the Belle Collaboration~\cite{Belle-18,Belle-21,Belle-19}.
In PDG, it has been nominated as a three-star resonance with $\Gamma= 6.4 ^{+3.0}_{-2.6}$~MeV, but its spin and parity quantum
numbers are yet to be measured~\cite{PDG20}.
In quark models, the mass and width of this state can be best explained by assuming an orbital excitation with 
$J^P=3/2^-$~\cite{XZ18,LWLZ19,WGLXZ18}.
Although the quark model provides a natural explanation, however, there exist molecular interpretations as its mass is close 
to the $\Xi(1530)\bar{K}$ threshold~\cite{PO18,Valderrama18, GL19, ITO20, LZWXG20}.

\begin{table}[b]
	\begin{ruledtabular}
		\renewcommand{\arraystretch}{1.6}
		\centering
		\caption{\label{omega2012} 
		Predicted decay width of $\Omega(2012)$ in the units of MeV. 
		We regard this resonance as the $\ket{70,{}^{2}10,1,1,\frac{3}{2}^-}$ state. 
		We use the mass of initial and final states from PDG~\cite{PDG20}.
		The branching fraction for each decay mode is indicated in parentheses. }
		\begin{tabular}{cccccc}
			State 				 			 & Channel 			   & $\Gamma_{\rm NR}$   & $\Gamma_{\rm NR+RC}$	&$\Gamma_{\rm Expt.}$~\cite{PDG20} \\ \hline
			$\Omega(2012)$	   & $\Xi \bar{K} $	  		  & 2.63 (96.7\%)				 			& 2.41 (95.6\%)\\
			& $\Xi \bar{K} \pi$		  & 0.09 (3.3\%)					  	 		 & 0.11 (4.4\%)	\\
			& Sum 			  		& 2.72				  			  & 2.52 				  	 & $6.4 ^{+3.0}_{-2.6}$\\ 
		\end{tabular}
		\renewcommand{\arraystretch}{1}
	\end{ruledtabular}
\end{table}

\begin{figure}[t]
	\centering
	\includegraphics[scale=0.4]{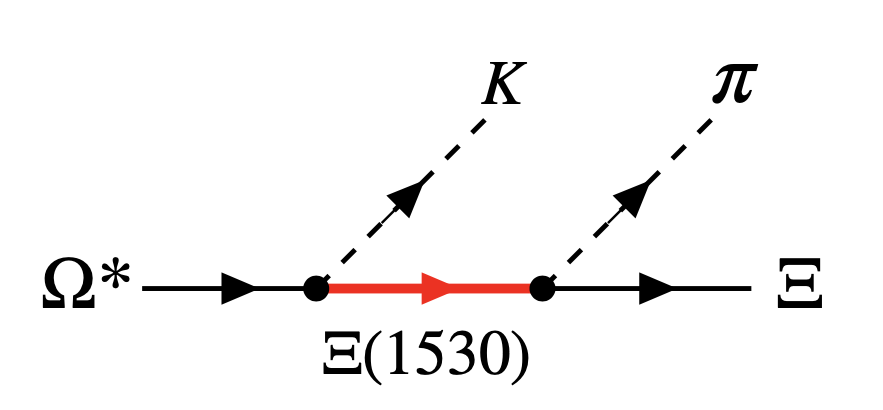}
	\caption{\label{3body} Three-body decay of $\Omega^* \to \Xi \bar{K}  \pi$ through $\Xi(1530)$. 
	We sum up the contributions from all possible channels. 
	We note that the intermediate states can be $\Xi(1530)^-$ and $\Xi(1530)^0$.}
\end{figure}

In the present work, we compute the decay width of $\Omega(2012)$ in the quark model including the relativistic corrections. 
In addition to the $\Xi \bar{K} $ channel, the three-body decay of $\Xi \bar{K} \pi$ is also considered.
Since the mass of $\Omega(2012)$ lies below the $\Xi(1530) \bar{K}$ threshold, $\Omega(2012) \to \Xi(1530) \bar{K}$ decay is not allowed. 
However, its decay into the three-body final state of $\Xi \bar{K} \pi$ is allowed but its decay width has not been estimated in 
any quark model calculation.
In the present work, we consider this decay as a sequential process going through $\Xi(1530)$ as shown in Fig.~\ref{3body}.
This process contains two interaction vertices, namely, $\Omega(2012) \to \Xi(1530)\bar{K} $ and $\Xi(1530)\to \Xi\pi$.%
\footnote{One may also consider two-step decay process of $\Omega(2012) \to \Xi \bar{K}^* \to \Xi (\bar{K} \pi)$.
However, the threshold energy of the $\Xi \bar{K}^* $ channel is about 2207~MeV which is much larger than the mass of $\Omega(2012)$.
On the other hand, the $\Xi(1530) \bar{K}$ channel has the threshold energy of about 2025~MeV.}
For the first vertex of $\Omega(2012) \to \Xi(1530)\bar{K} $, we use the interaction Hamiltonian with and without 
relativistic corrections.
However, for the second vertex of $\Xi(1530)\to \Xi\pi$, we extract the coupling of the quark model with the relativistic corrections
to reproduce the data as shown in Table~\ref{tab:xi1530}.
The details of the three-body decay calculation are given in Appendix.

\begin{figure}[t]
	\centering
	\includegraphics[scale=0.18]{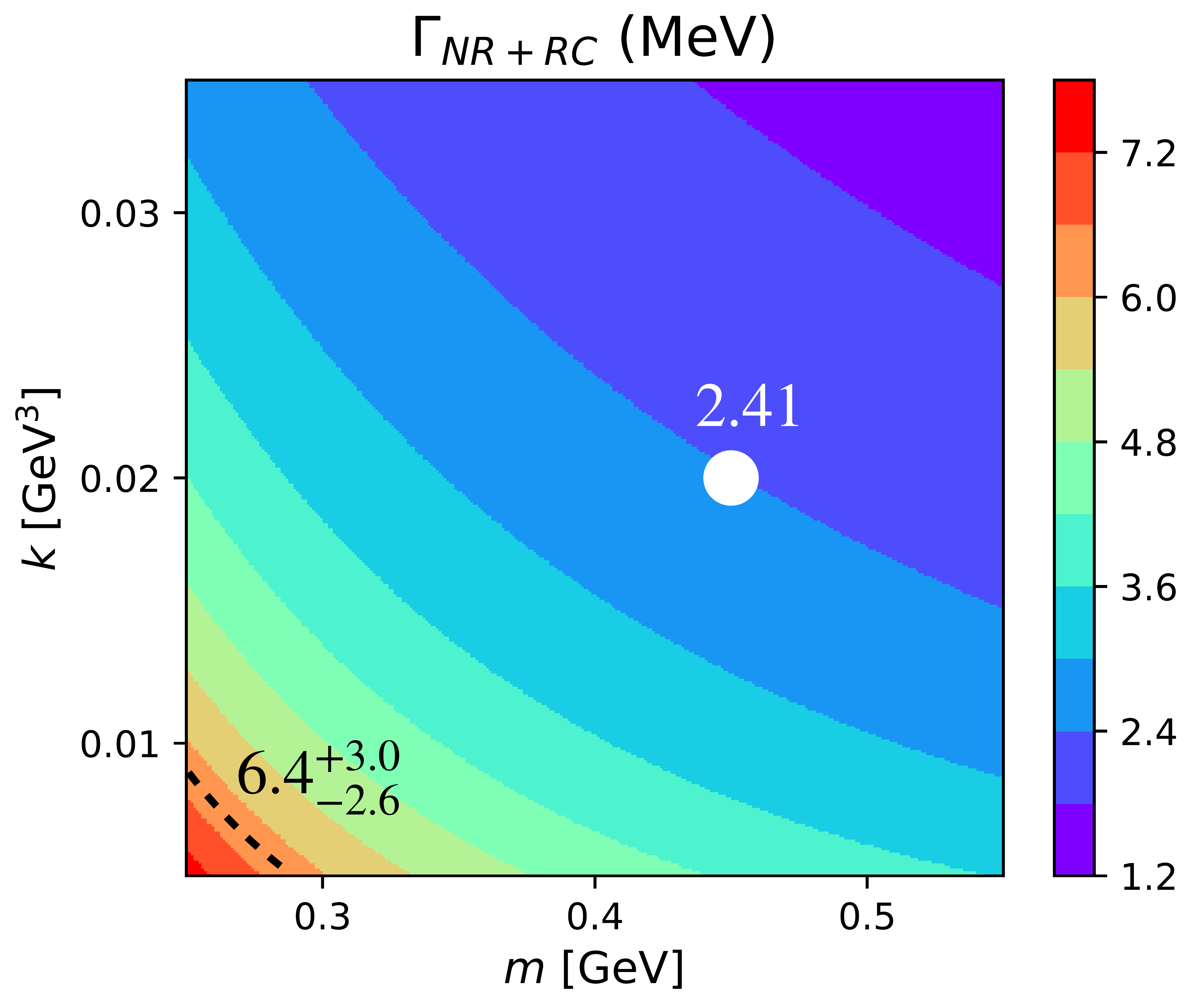}
	\caption{\label{o2012} 
	Contour plot of the decay width of $\Omega(2012)\to \Xi \bar{K}$ with the $\ket{70,{}^{2}10,1,1,\tfrac{3}{2}^-}$ configuration 
	as a function of the quark mass $m$ and spring constant $k$.}
\end{figure}

Our results for the decays of $\Omega(2012)$ as a $J^P=3/2^-$ state are given in Table~\ref{omega2012}.
The predicted width is at the order of a few MeV, which is in fair agreement with the data, $\Gamma = 6.4 ^{+3.0}_{-2.6}$~MeV.
The contour plot for the decay width as a function of the model parameters is shown in Fig.~\ref{o2012}.
If $\Omega(2012)$ is regarded as a $J^P=1/2^-$ state, the predicted width becomes much larger being around 20~MeV 
because of the $s$-wave nature of the $\Xi \bar{K} $ decaying channel.
In this case, the obtained decay width overestimates the experimental value.
Therefore, $\Omega(2012)$ is most likely to be a $J^P=3/2^-$ state, which supports the conclusion of other quark model 
calculations reported in Refs.~\cite{XZ18,WGLXZ18,LWLZ19}.

\begin{figure}[t]
	\centering
	\includegraphics[scale=0.2]{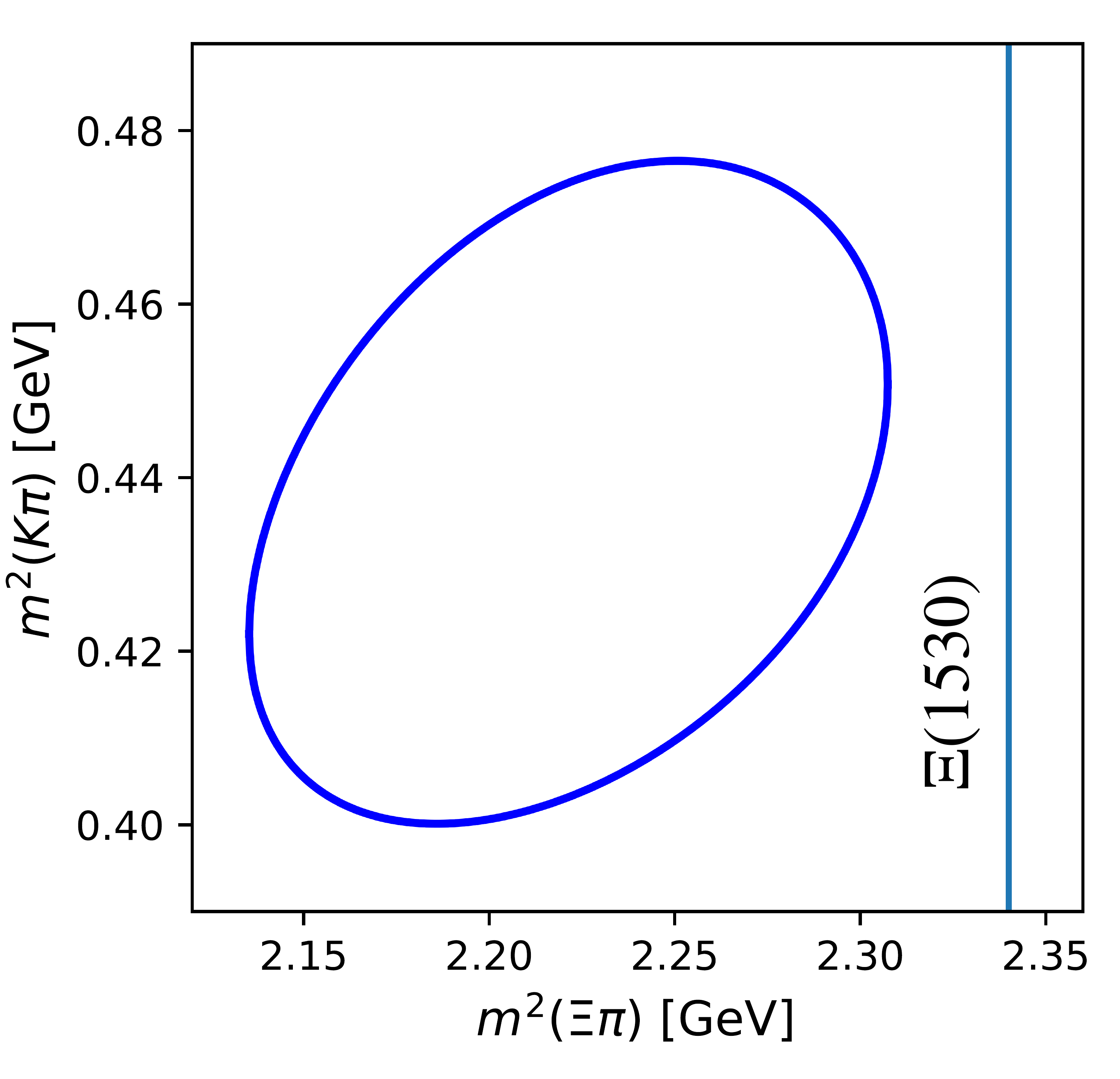}
	\caption{\label{dalitz_plot} 
	The Dalitz plot of $\Omega^* \to \Xi \bar{K}  \pi$ decay. 
	The $\Xi(1530)$ resonance band appears outside the Dalitz plot because of the limited phase space.}
\end{figure}

In addition, the width of the $\Omega(2012) \to \Xi \bar{K} \pi$ decay is also  tabulated in Table~\ref{omega2012}.
We find that this partial decay width is suppressed although $\Omega(2012)$ decays into the $\Xi(1530)\bar{K} $ channel in $s$ wave.
It can be understood by the fact that $\Xi(1530)$ resonance band is located well outside the Dalitz plot as shown in Fig~\ref{dalitz_plot}.
As for the comparison with the recent data measured by the Belle Collaboration~\cite{Belle-19}, we compute the ratio defined by
\begin{eqnarray}
	R^{\Xi \pi \bar{K}}_{ \Xi \bar{K} } &=& \frac{\Gamma(\Omega^*\to \Xi(1530) \bar{K}  \to \Xi \pi \bar{K}) }{\Gamma (\Omega^*\to \Xi \bar{K} ) }.
\end{eqnarray}
With the relativistic corrections included, the ratio is estimated by the quark model as
\begin{eqnarray}
	\left[ R^{\Xi \pi \bar{K}}_{ \Xi \bar{K} } \right]_{\rm QM} &\approx&  4.5\%,
\end{eqnarray}
which is consistent with the measured ratio~\cite{Belle-19}
\begin{eqnarray}
	\left[ R^{\Xi \pi \bar{K}}_{ \Xi \bar{K} } \right]_{\rm expt} < 11.9 \%.
\end{eqnarray}
It is worth noting that no significant signal of $\Omega(2012)$ is found in the experimentally measured $\Xi \bar{K} \pi$ invariant mass 
and only the upper bound of the branching ratio is provided.
In this regard, we provide a further evidence that the $\Omega(2012)$ is well explained in the quark model as a $J^P=3/2^-$ state.
This contrasts with the assumption of molecular structure of $\Omega(2012)$.
In this case, a large value of the ratio $R^{\Xi \bar{K} \pi}_{\Xi \bar{K} }$ is already predicted ~\cite{PO18,Valderrama18},
which is not consistent with the observed ratio although further refinements seem to make it difficult to rule out the interpretation
of molecular structure~\cite{GL19, ITO20, LZWXG20}. 

\begin{table}[t]
	\begin{ruledtabular}
		\renewcommand{\arraystretch}{1.6}
		\centering
		\caption{\label{omega1957} 
		Predicted decay width of the LS partner of $\Omega(2012)$ with the $\ket{70,{}^{2}10,1,1,\frac{1}{2}^-}$ configuration
		in the units of MeV. Its mass is assumed to be 1957 MeV~\cite{LWLZ19}.
		}
		\begin{tabular}{ccccc}
			State 				  			  					& Channel 				  & $\Gamma_{\rm NR}$  & $\Gamma_{\rm NR+RC}$ \\ \hline
			$\ket{70,{}^{2}10,1,1,\frac{1}{2}^-}$	 	&	  	 $\Xi \bar{K}$	 	    & 17.5 			    			& 19.5	\\ 
		\end{tabular}
		\renewcommand{\arraystretch}{1}
	\end{ruledtabular}
\end{table}

\begin{figure}[t]
	\centering
	\includegraphics[scale=0.2]{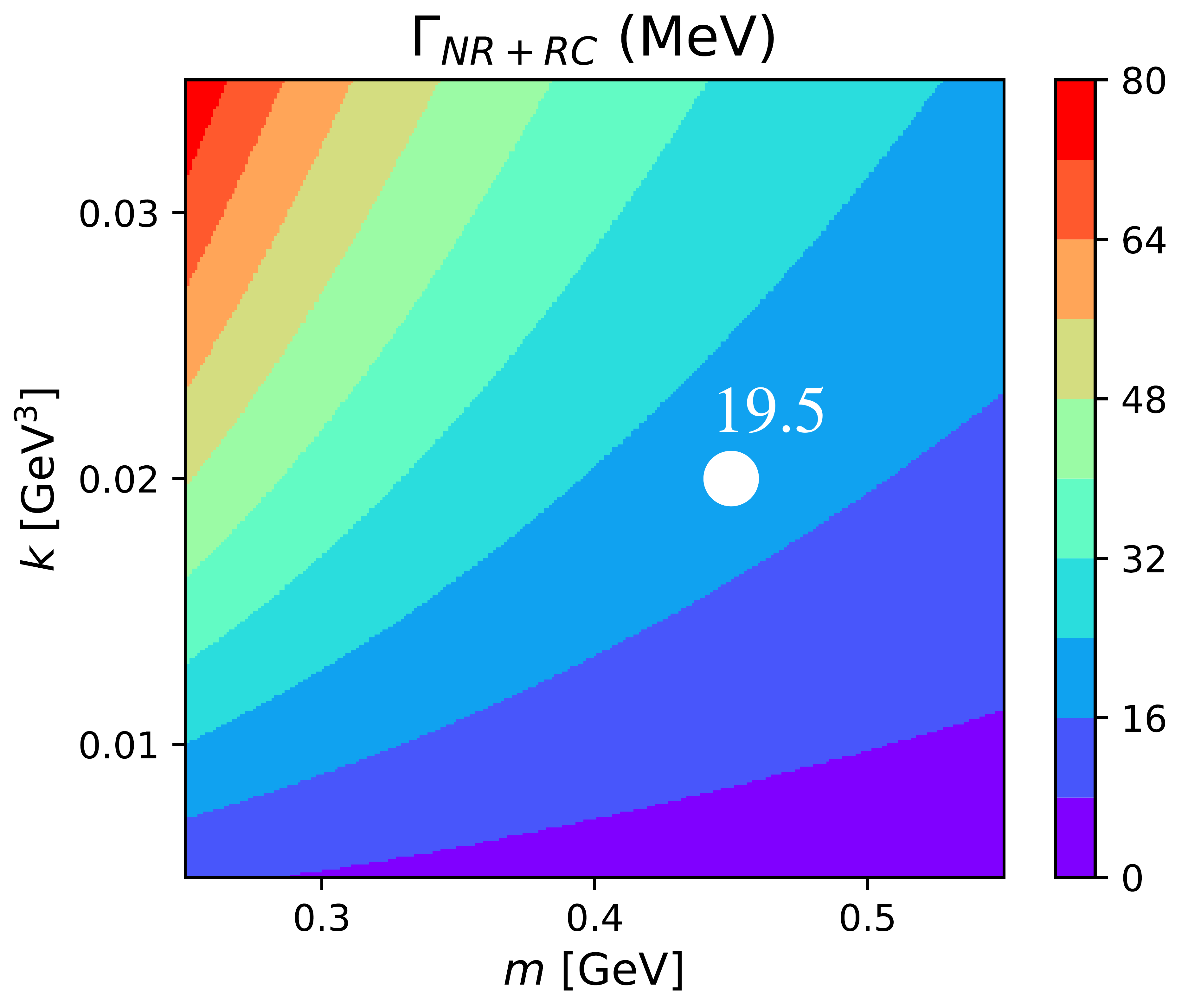}
	\caption{\label{o1950} 
	Contour plot of the decay width of the LS partner of $\Omega(2012)$ with the $\ket{70,{}^{2}10,1,1,\tfrac{1}{2}^-}$ 
	configuration as a function of the quark mass $m$ and spring constant $k$. Its mass is assumed to be 1957~MeV~\cite{LWLZ19}.}
\end{figure}

\subsection{\boldmath LS partner of the ${\Omega(2012)}$}

To further understand the internal structure of $\Omega(2012)$, more precise measurements of the decay width and 
determination of its spin and parity are certainly needed.
Another clue to understand its structure would be found by searching for its spin-orbit (LS) partner of $J^P=1/2^-$.
In the quark model, it is natural to expect to have its LS partner, while it is hard to generate such a state
in the molecular scenario of $\Omega(2012)$.
Therefore, the discovery of the LS partner would a useful tool to distinguish the two models.

In the quark model with SU(6) classification, there are two negative parity $\Omega$ baryons, i.e., the 
$\ket{70,{}^{2}10,1,1,\frac{1}{2}^-}$ and $\ket{70,{}^{2}10,1,1,\frac{3}{2}^-}$ states~\cite{CI86,LWLZ19}.
If the quark model picture is justified and supported, the $1/2^-$ state should be discovered.
Furthermore, precise measurements of their masses would shed light on our understanding of the vanishing LS splitting 
in the light baryon sector~\cite{OT89}.

In this exploratory study, we adopt the mass of the $1/2^-$ state as 1957~MeV following the quark model of Ref.~\cite{LWLZ19} and 
compute its decay width.
With this mass, only the decay channel into $\Xi \bar{K}$ is open and the results are shown in Table~\ref{omega1957}.
Figure~\ref{o1950} shows the parameter dependence of the decay width calculated in our model.
In contrast to the case of $\Omega(2012)$ of $J^P=3/2^-$, this shows that the predicted width for the $1/2^-$ state has 
large model uncertainties.  
However, a width of around 20~MeV is obtained by using the central values of the parameters and we find that
the obtained width is insensitive to the variation of the initial mass.
Thus this missing resonance could be searched for in current experimental facilities.

\begin{table}[t]
	\begin{ruledtabular}
		\renewcommand{\arraystretch}{1.6}
		\centering
		\caption{\label{omega2159} 
		Predicted decay width of the Roper-like $\Omega$ resonance with the $\ket{56,{}^{4}10,2,0,\frac{3}{2}^+}$ configuration
		in the units of MeV. 
		Its mass is assumed to be 2159 MeV~\cite{LWLZ19}. 
		The branching fraction for each decay mode is indicated in parentheses.}
		\begin{tabular}{ccccc}
			State 				  			 &  Channel 		  	& $\Gamma_{\rm NR}$ & $\Gamma_{\rm NR+RC}$ \\ \hline
			$\ket{56,^{4}10,2,0,\frac{3}{2}^+}$	  	   & $\Xi \bar{K} $				& 8.4	(38.7\%)			  &  68.2  	(58.2\%) \\
			& $\Xi(1530) \bar{K} $					& 13.3		(61.3\%)		  		& 48.9 	(41.8\%) \\
			& sum 			  		  	  & 21.7			    		& 117 \\ 
		\end{tabular}	
		\renewcommand{\arraystretch}{1}
	\end{ruledtabular}
\end{table}

\begin{figure}[t]
	\centering
	\includegraphics[scale=0.18]{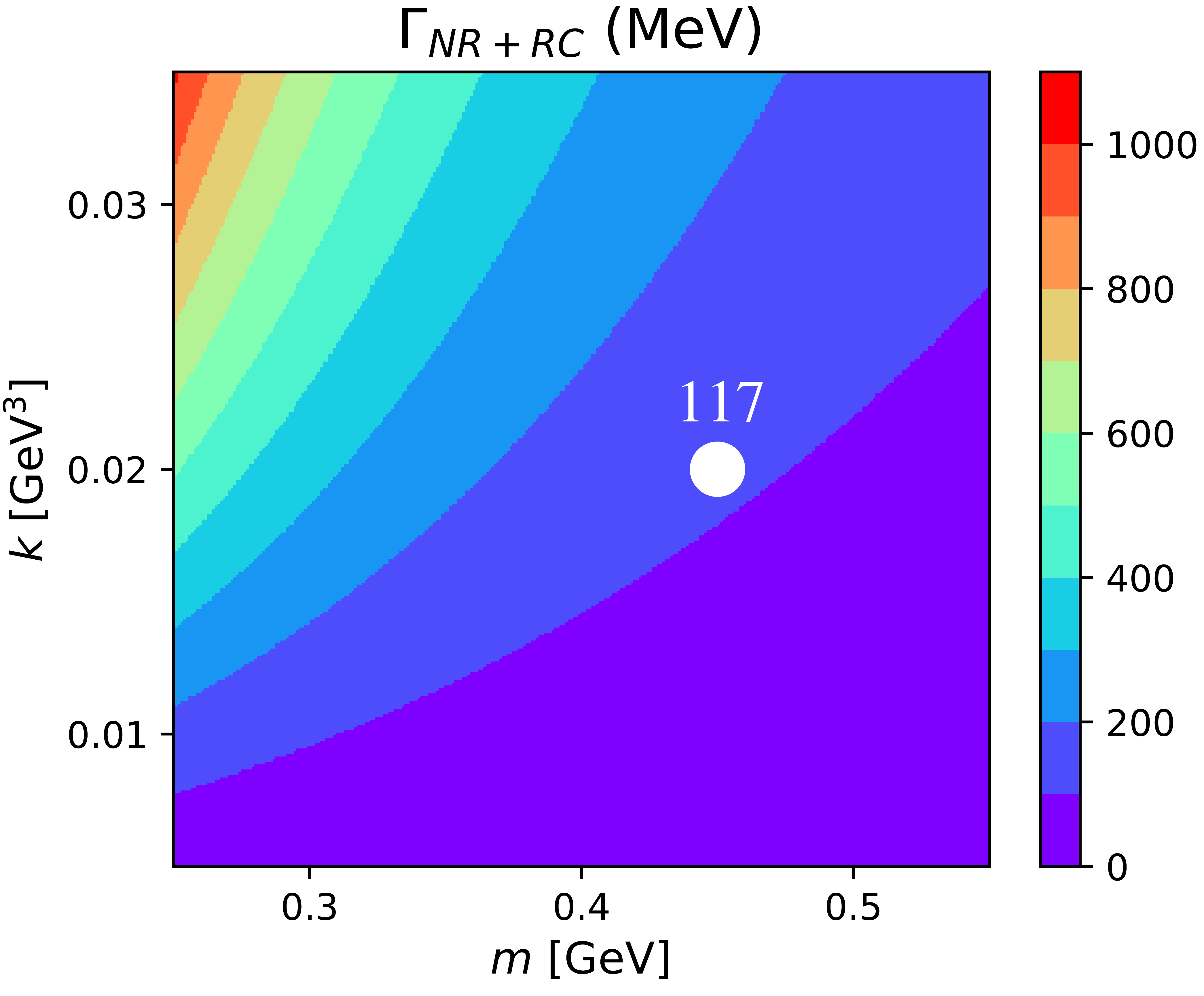}
	\caption{\label{o2159} 
	Contour plot of the decay width of the Roper-like $\Omega$ resonance with the $\ket{56,{}^{4}10,2,0,\frac{3}{2}^+}$ configuration
	as a function of the quark mass $m$ and spring constant $k$. Its mass is assumed to be 2159 MeV~\cite{LWLZ19}.}
\end{figure}

\subsection{\boldmath Roper-like ${\Omega}$ baryon }

As in the $\Xi$ spectrum, we now consider the decay of the Roper-like $\Omega$ resonance in the quark model.
The first radial excitation of the $\Omega$ baryon has $J^P=3/2^+$ with the configuration of $\ket{56,{}^{4}10,2,0,\frac{3}{2}^+}$.
Several quark model calculations predict that its decay width is rather small~\cite{XZ18,LWLZ19}.
However, as we have learned in our previous studies, the quark model with the nonrelativistic interactions in Eq.~(\ref{nonrel}) 
failed to reproduce the observed large width of Roper-like heavy baryons~\cite{ANHT20a} and the relativistic effects are found 
to play an important role, in particular, in the decays of radially excited states~\cite{ASH21}.
Therefore, it would be meaningful to see how the relativistic corrections contribute to the decay width of the Roper-like $\Omega$ baryon.

The Roper-like resonances are expected to have a mass about 400-500~MeV larger than the ground states.
In our model, the radially excited states of the octet baryons and anti-triplet heavy baryons are discussed in Ref.~\cite{ANHT20b}.
Here we extend our work to the radially excited states of decuplet baryons as well.
The prediction on the mass of this state is model-dependent~\cite{MR21b,  LWLZ19,CIK81} and we take its mass as 2169~MeV
following the quark model calculation of Ref.~\cite{LWLZ19}.
The obtained width with two open channels is shown in Table~\ref{omega2159} and its contour plot is given in Fig.~\ref{o2159}.
As expected, the relativistic corrections are quite large similar to the cases of other siblings studied in Ref.~\cite{ASH21}.
The width is estimated to be around 100~MeV although large model uncertainties are seen as depicted in Fig.~\ref{o2159}.
Therefore, experimental search will help understand the internal structure of the Roper resonance and its analog states.

\section{Summary} \label{sec:summary}

In the present work, we have investigated the decays of the excited states of $\Xi$ and $\Omega$ baryons.
In particular, we focus on the possible Roper-like states in these baryon spectra.
The data collected in Fig.~\ref{roper} show that the Roper resonance and its siblings have the excitation energy close to 500~MeV 
from the corresponding ground state,
which leads to the speculation that there might exist such analogous states with a similar energy gap in multi-strangeness baryons as well.
In fact, there are a few states observed in $\Xi$ baryons in this mass region, and they are classified as a single resonance as explained 
in PDG~\cite{PDG20}.
Therefore, more precise measurements in various reactions at current and planned experimental facilities~\cite{KLF-20,PANDA-21,JPARC-21} 
will clarify the issue and will provide a decisive information on the structure of radially excited states.

Being motivated by this observation, we have explored the decays of the low-lying $\Xi$ and $\Omega$ resonances in various decay
channels including the relativistic corrections.
As in the case of heavy baryons~\cite{ASH21}, the relativistic corrections generally play an important role in the decays of decuplet 
states and the Roper-like states.
In the case of negative parity states, however, such corrections are found to be insignificant.

In the quark model, we found that the decay properties of $\Xi(1620)$ and $\Xi(1690)$ resonances favor 
the spin-parity quantum numbers of $J^P=1/2^-$.
Assumption of $J^P = 1/2^+$ for these states results in small decay widths, which is very difficult to explain the data.
For the $\Xi(1820)$ resonance, we found that the decay width is compatible with the $3/2^-$ assignment~\cite{TDDG78,BBBB87b}.
But, inconsistencies among the data on its decay width and branching ratios should be further clarified in future experiment as
there might be a Roper-like state with a similar mass.
We also found that if $\Xi(1950)$ is assigned as a $J^P=1/2^+$ state, the predicted width becomes much larger than the data and 
its excitation energy is quite large as compared to other Roper-like resonances. 
However, we expect that the radial excitation of the decuplet $\Xi$ baryon with $J^P=3/2^+$ may exist in this energy region.
The predicted decay properties of this Roper-like state would be useful to search for this state in experiments.

As for $\Omega$ baryons, we found that the newly observed $\Omega(2012)$ could be naturally explained as a $J^P=3/2^-$ state 
in the quark model.
In particular, we provide another evidence; namely, the quark model prediction is consistent with the ratio 
$R^{\Xi \bar{K} \pi}_{\Xi \bar{K} }$ observed by the Belle Collaboration~\cite{Belle-19}.
As another test of the quark model, we suggest searching for the LS partner of $\Omega(2012)$ since its existence may distinguish 
the quark model and molecular pictures.
Discovery of this state would be crucial to further clarify its internal structure.

For the Roper-like multi-strangeness resonances, we expect that they are broad resonances having decay widths of larger than 100~MeV. 
This is very different from the predictions of the leading contribution of
nonrelativistic quark models~\cite{XZ18,LWLZ19} because of the crucial role
of the relativistic corrections.
In fact, other Roper-like resonances such as $\Lambda(1600)$ and $\Sigma(1660)$ have large decay widths of around~200 MeV.
It is worth noting that a large branching fraction to $\Sigma \bar{K}$ and $R^{\Xi\pi}_{\Xi(1530)\pi}$ close to unity can be a signature of the Roper-like $\Xi$ resonance with $J^P=1/2^+$.
Further experimental investigation of a broad resonance in the excitation energy of around 500~MeV is highly desirable to find the missing 
Roper-like multi-strangeness resonances. 
Although precise understanding of the structure of these resonances requires the inclusion of the meson cloud effects, our
findings will serve as a guide for a more sophisticated and refined theoretical approaches.

\acknowledgments

A.J.A was supported by the YST Program at the APCTP through the Science and Technology Promotion Fund and Lottery Fund of the Korean Government and also by the Korean Local Governments - Gyeongsangbuk-do Province and Pohang City.
He is also grateful to the organizers of ``Workshop on physics at the J-PARC K10 beam line'' that motivated the present work.
A.H. was supported in part by Grants-in Aid for Scientific Research on Innovative Areas (No. 18H05407).
The work of Y.O. was supported by the National Research Foundation of Korea (NRF) under Grants
No. NRF-2020R1A2C1007597 and No. NRF-2018R1A6A1A06024970 (Basic Science Research Program).
Y.O. also acknowledges the support from the APCTP Senior Advisory Group.

\appendix*

\section{Three-body decay of $\mathbf{\Omega(2012)}$}

In this Appendix, we present the details of computing the three-body decay of the $\Omega(2012)$ with $J^P=3/2^-$ into $\Xi \bar{K} \pi$. 
We consider this decay as the sequential process going through the $\Xi(1530)$ in the intermediate state as depicted in Fig.~\ref{3body}.
In such a decay, the $\Xi(1530)$ contributes virtually to the $\Xi \bar{K} \pi$ decay because of the limited phase space as illustrated in Fig.~\ref{dalitz_plot}.
However, due to the $s$-wave nature of the $\Xi(1530)\bar{K}$ system, the tail contribution is expected to be non-negligible.
This situation is similar to the decay of $\Lambda_c(2625) \to\Lambda_c\pi\pi$ where $\Sigma_c(2525)$ is off-shell~\cite{ANH17}.
On top of this decay mechanism, one may also consider the contribution from the $\Xi K^*$ intermediate state from which the $K^*$ meson 
decays into the $\bar{K} \pi$ system.
But, it is highly off-shell so that its contribution is expected to be suppressed and will not be considered in the present calculation.

In order to describe the decay process, we employ the effective Lagrangian in nonrelativistic approximation to model the three-body decay as 
a successive two-body decays. 
The first vertex, $\Omega(2012) \to \Xi(1530)\bar{K}$, in Fig.~\ref{3body} is the decay of $\textstyle \frac32^- \to \frac32^+ + 0^-$
and its amplitude is given by the following two vertices,
\begin{subequations}
	\label{app:eqn1}
	\begin{eqnarray}
		-i\mathcal{T}_{\Omega^* \to \Xi^*\bar{K} }^{(s)}  &=& g_{1}^s \chi^\dagger_{\Xi^*} \chi_{\Omega}^{}, \\
		-i\mathcal{T}_{\Omega^* \to \Xi^*\bar{K} }^{(d)} 	&=& g_{1}^d \chi^\dagger_{\Xi^*} 
		\left( V_{ij}\ q_{1i}^{} q_{1j}^{} \right) \chi_{\Omega}^{},
	\end{eqnarray}
\end{subequations}
for $s$ and $d$ wave channel, respectively, where $V_{ij}$ represents the spin transition operator from $j=3/2$ to $j=3/2$ in 
$d$ wave~\cite{ANHT20a}.
The spinors of the $\Omega(2012)$ and $\Xi(1530)$ are represented by $\chi_{\Omega^{*}}$ and $\chi_{\Xi^*}$, respectively.
The outgoing kaon momentum is denoted by $q_1^{}$.

The second vertex, which describes the $\Xi(1530) \to \Xi\pi$ decay, has only the $p$-wave component and is written as
\begin{eqnarray}
	-i\mathcal{T}_{\Xi^* \to \Xi \pi}^{} &=& g_{2}^{p} \chi^\dagger_{\Xi} \left( {\bf S} \cdot {\bf q}_2^{} \right)   \chi_{\Xi^*}^{}, 	
	\label{app:eqn2}
\end{eqnarray}
where the spin transition operator ${\bf S}$ is introduced for the transition from $j=3/2$ to $j=1/2$, $q_2^{}$ is the outgoing pion momentum, 
and $\chi_\Xi^{}$ is the spinor of the $\Xi$ baryon.
The coupling strengths $g_1^s$, $g_1^d$, and $g_2^p$ are determined by the quark model by using the helicity amplitudes of these decays.
The detailed description can be found, for example, in Ref.~\cite{ANHT20a} and will not be repeated here.

The three-body decay amplitude of $\Omega(2012)\rightarrow \Xi \bar{K}  \pi $ through the intermediate $\Xi(1530)$ as described 
in Fig.~\ref{3body} is then obtained as
\begin{eqnarray}
	-i\mathcal{T} = - i \frac{\mathcal{T}_{\Xi(1530) \rightarrow \Xi\pi} \mathcal{T}_{\Omega \rightarrow \Xi(1530) \bar{K} } }
	{m_{\Xi\pi} - M_{\Xi(1530)}+\frac{i}{2}\Gamma_{\Xi(1530)}}, 
\label{amp1}
\end{eqnarray}
using the expressions of Eqs.~(\ref{app:eqn1}) and (\ref{app:eqn2}), where $m_{\Xi\pi}$ is the invariant mass of the $\Xi\pi$ system, 
and $M_{\Xi(1530)}$ and $\Gamma_{\Xi(1530)}$ are the mass and width of the $\Xi(1530)$, respectively,
whose values are taken from PDG~\cite{PDG20}.
Then, we need to sum up all possible channels.
The decay width is then computed as 
\begin{eqnarray}
	\Gamma = \frac{1}{(2\pi)^3}\frac{1}{32M^3_i} \int   \overline{|\mathcal{T} |^2} \, dm^2_{\Xi \pi}  \, dm^2_{\pi \bar{K} }, 
	\label{dalitz}
\end{eqnarray}
where $M_i$ is the initial mass of $\Omega(2012)$.

\end{document}